\documentclass[12pt,letterpaper]{article}
\usepackage[a4paper, total={6.5in, 9in}]{geometry}

\usepackage{graphicx}
\usepackage[table]{xcolor}
\usepackage{helvet}
\usepackage{authblk}
\usepackage{hyperref}
\usepackage{amsmath} 
\usepackage{amssymb} 
\usepackage{orcidlink} 
\usepackage{booktabs,makecell}
\usepackage{floatrow}
\usepackage{natbib}
\usepackage{verbatim}
\usepackage{mathtools}
\usepackage{amsthm}
\usepackage{caption}
\usepackage{bm}

\theoremstyle{plain}
\newtheorem{theorem}{Theorem}

\newtheorem{remark}{Remark}

\makeatletter
\renewcommand{\maketitle}{\bgroup\setlength{\parindent}{0pt}
\begin{flushleft}
  \textbf{\Large \@title}
  
  \@author
\end{flushleft}\egroup}
\makeatother

\title{Contamination dynamics of active rods in long microchannels}
\date{}

\author[1,2]{Jason Palos}
\author[2]{Chase Brown}
\author[2,*]{Mykhailo Potomkin}
\author[3]{Shawn D. Ryan}

\affil[1]{School of Mathematics, University of Minnesota, Minneapolis, Minnesota 55442, USA}

\affil[2]{Department of Mathematics, University of California, Riverside, 900 University Avenue Riverside, California 92521, USA}

\affil[3]{Department of Mathematics and Statistics and Center for Applied Data Analysis and Modeling, Cleveland State University, Cleveland, Ohio 44115, USA}

\affil[*]{corresponding author}

\begin{document}
\maketitle

\begin{abstract}
   Active rods are elongated, self-propelled particles that serve as effective models for motile microorganisms, such as bacteria and sperm cells. When confined to a microfluidic channel, active rods exhibit a distinctive motility pattern: they tend to swim against the fluid flow along channel walls. In pressure-driven channel flow, they alternate between long upstream runs along the walls and short downstream spurts in the bulk. This behavior has practical implications for bacterial contamination of catheters and microfluidic devices against the flow. We study the spreading of a population of active rods along long microchannels using a hierarchy of models.
Monte Carlo simulations of a three-dimensional model show
that net upstream transport depends non-monotonically on the flow rate,
decreases with rotational diffusion, and depends on the shape of the channel
cross-section. The simulations also show that upstream and downstream
persistence times follow exponentially decaying distributions. This motivates
the derivation of a one-dimensional, two-state velocity-jump model. For
exponentially distributed persistence times, we derive the evolution equation
for the concentration of continuously injected rods and determine its long-time
limit in closed form. This analysis reveals a concentration plateau upstream of
the source when the mean upstream velocity is positive and exponential decay
otherwise. Using parameters measured from the three-dimensional simulations,
the one-dimensional model reproduces the concentration profiles from the Monte Carlo
simulations, and these results are consistent with recent experimental
observations of bacterial contamination in microfluidic devices.

\end{abstract}


\section{Introduction}


Active matter systems consist of many motile agents capable of self-propulsion. Examples include biological micro-organisms such as motile bacteria and sperm cells \cite{aranson2022bacterial, rothschild1963non}, as well as synthetic materials such as bimetallic rods \cite{paxton2005motility,paxton2006chemical}. Such systems are out-of-equilibrium due to the constant input of energy \cite{hallatschek2023proliferating,bar2020self}, resulting in numerous effects such as migration against fluid flow \cite{hill2007hydrodynamic, marcos2012bacterial, palacci2015artificial, yuan2015hydrodynamic, ren2017rheotaxis} and viscosity reduction \cite{sokolov2009reduction,haines2009three,ryan2011viscosity, potomkin2016effective}. One effective model for elongated particles within active matter systems are active rods -- finite segments that are self-propelled in the direction they are facing \cite{bar2020self, potomkin2017focusing, rubio2021self}. 

Two key characteristics of active rods swimming in a confined environment are bordertaxis, the tendency for rod accumulation at the boundary, and negative rheotaxis, the ability of active rods to swim upstream against a fluid flow. Bordertaxis can arise via mechanisms such as hydrodynamic attraction to nearby surfaces \cite{drescher2011fluid}, but it does not necessarily require a background fluid flow. Instead, bordertaxis emerges primarily from the active nature of rods' motion: a confined active rod will eventually collide with and follow the boundary until it escapes via some mechanism such as random reorientation or torque exerted by the wall \cite{brown2025boundary,berlyand2026multiscale}. On the other hand, negative rheotaxis occurs via reorientation by the background flow. Due to the no-slip boundary condition, the end of an active rod closest to the wall will experience lower drag by the background flow compared to the end nearest to the channel centerline. This imbalance of forces being exerted on the rod generates a torque that reorients it to face against the direction of the flow, at which point the rod can propel itself upstream. Bordertaxis and negative rheotaxis, together with rotational diffusion, result in typical trajectories where active rods alternate between intervals of upstream swimming near the boundary and downstream swimming in the fluid bulk. These two phenomena and some statistical properties of their resulting trajectories have been studied mathematically in various geometries, such as nozzles \cite{potomkin2017focusing, rubio2021self}, chambers \cite{ezhilan2015transport}, and elliptic microchannels \cite{brown2025boundary}. For active Brownian particles in pressure-driven channel flow, the effective drift and longitudinal dispersion have been derived by a generalized Taylor dispersion analysis \cite{peng2020upstream}. 

Active rods' ability to exhibit negative rheotaxis and swim against a background flow possesses important implications in certain biomedical contexts. For instance, the presence of motile bacteria in ducts where a fluid flows in a single direction, such as urine catheters and the urinary tract, can result in upstream swimming-induced bacterial infections \cite{wright2005uropathogenic,dohnt2011vitro}. Such infections are highly prevalent in health care institutions, where an estimated 70-80$\%$ of all bacterial urinary tract infections are attributed to urine catheter devices \cite{nicolle2014catheter}. Additionally, recent experiments \cite{figueroa2020coli} report that in spite of their run-and-tumble dynamics, \textit{Escherichia coli} bacteria can travel distances of $\sim$15mm toward a fluid source in as little as 15 minutes by swimming upstream near the corners and surfaces of a rectangular microfluidic channel. This observation was termed \textit{super-contamination} for the ability of the bacterial population's front to persist upstream in the microchannel at velocities nearly independent of the corresponding fluid flow rate. Because this super-contamination effect was linked to upstream swimming near corners, surfaces, and crevices of a microchannel, the cross-sectional geometry of the channel and regions of high or low curvature may significantly influence upstream transport dynamics. Indeed, in microchannels of rectangular cross-section, rods tend to accumulate and migrate at corners resulting in four lanes of transport \cite{hill2007hydrodynamic, figueroa2015living, bukatin2020self, brown2025boundary}. Moreover, identifying any geometric controls of the upstream transport of active rods may provide insight into mitigation of upstream bacterial infections. Recent experiments support this view: upstream invasion of microstructured devices by {\it E. coli} depends on channel geometry and flow strength \cite{tao2026invasion}, and catheters whose inner-wall geometry was optimized to disrupt upstream swimming reduced upstream contamination by one to two orders of magnitude \cite{zhou2024ai}.


Besides its pertinence to biomedicine, the run-and-tumble motility pattern of swimming bacteria has emerged as a model system in active matter and out-of-equilibrium statistical physics, motivating experimental and theoretical studies of collective motion \cite{dombrowski2004self, marchetti2013hydrodynamics}, enhanced and anomalous diffusion \cite{olsen2024optimal, thiel2012anomalous, pietrangeli2024optimal, villa2020run, malakar2018steady, grossmann2016diffusion}, cellular transport and chemotaxis \cite{mallikarjun2023chiral, rivero1989transport,ahmed2008experimental}, as well as migration up concentration gradients \cite{galajda2007wall}. Stemming from measurements of rotational switching statistics of \textit{E. coli}'s flagellar motor \cite{berg1972chemotaxis,berg2004coli}, most modeling and theoretical studies have historically assumed that run-and-tumble dynamics follow a Poisson process and constitutes an isotropic random walk with exponentially distributed durations of run and tumble phases \cite{berg1993random,lovely1975statistical,codling2008random,othmer1988models,aranson2022bacterial, zaburdaev2015levy}. However, a heavy-tailed power law distribution has also been observed to describe the durations of counterclockwise rotation of the flagellar motor that drives bacteria forward during the ``run'' phase \cite{korobkova2004molecular}. Similarly, a stochastic model for bacterial trajectories showed concordance with experimental spatiotemporal concentration profiles when the simulated run-times were described by a broad power-law distribution \cite{figueroa2020coli}. On the other hand, the model failed to qualitatively reproduce the experimental concentration profiles when the simulated run-times followed an exponential distribution, suggesting that the run-and-tumble statistics of singular bacteria may influence bacterial transport at the population level. Yet, the extent to which microscopic statistics such as rotation of the flagellar motor can affect the macroscopic properties of bacterial transport and dispersion remains the subject of ongoing research \cite{figueroa2020coli, sneddon2012stochastic, figueroa20203d, matthaus2011origin}.  

In this paper, we employ a hierarchy of mathematical models to elucidate the dynamics of active rods across different scales. Namely, our focus is on the dynamics of the active-rod concentration profiles along the major axis of the microchannel for two typical scenarios: (i) active rods are initialized randomly at the center of the channel, $z=0$, such that the subsequent concentration-profile dynamics resemble the fundamental solution of the advection-diffusion equation, when the $\delta$-function is taken as the initial condition; and (ii) active rods are injected at a constant rate at $z=0$, corresponding, for example, to the setting considered in \cite{figueroa2020coli}, where contamination due to the upstream swimming of bacteria was studied.  First, we utilize a three-dimensional model for active rods swimming inside a microfluidic channel to investigate the mechanisms facilitating their upstream transport. We demonstrate how statistical properties of active rods' trajectories depend on central physical parameters, such as the fluid flow rate and rotational diffusion. By examining simulated spatiotemporal concentration profiles and active rods' persistence time distributions in channels of both elliptical and rectangular cross-section, we reveal potential geometric controls of the upstream contamination process. Next, numerical simulations of the active rod model demonstrate that the essential transport dynamics can be isolated along the channel's major axis and described by an effective one-dimensional biased random walk process. By considering the one-dimensional mathematical model for rod transport, we examine how the macroscopic upstream contamination of the active rod population depends on statistical properties of the underlying distributions for active rods' persistence times. Simulation results from these one-dimensional models demonstrate that quantitatively distinct probability distributions governing individual rod persistence times can give rise to similar qualitative macroscopic contamination dynamics of the active rod population. Finally, we rigorously derive the partial differential equation governing the concentration profile in the one-dimensional model and obtain an explicit formula for its long-time behavior, consistent with the results of the three-dimensional simulations.

\section{Models}\label{sec:2}


\subsection{A three-dimensional model of active rods in a cylindrical microchannel}\label{subsec:3d-model}
To simulate active rod dynamics in a three-dimensional channel we recall the model we previously developed in \cite{brown2025boundary}. Its main simplifying assumptions are $(i)$  active rods do not interact with each other, corresponding to individual swimming or a dilute suspension of active rods, and $(ii)$ although active rods are advected and reoriented by the background flow, they do not perturb it. The state of an active rod is uniquely determined by its center location $\boldsymbol{r}$ and unit orientation vector $\boldsymbol{p}$. As time evolves, active rod's location and orientation satisfy the following system of equations: 
    \begin{equation}\label{eq:modelEquations}
    \left\{
    \begin{array}{l}
    \dot{\boldsymbol{r}} = V_0\boldsymbol{p} + \boldsymbol{u},\\
    \dot{\boldsymbol{p}} = (\text{I}-\boldsymbol{p}\boldsymbol{p}^{\text{T}})\left[\nabla \boldsymbol{u}\boldsymbol{p} + \sqrt{2D_{\text{rot}}}\,\dot{W}\right].
    \end{array}
    \right.
    \end{equation}
    These equations read as follows: an active rod's velocity is the sum of the propulsion with speed $V_0$ in the front direction, given by the orientation vector $\boldsymbol{p}$, and background flow $\boldsymbol{u}$ at the point $\boldsymbol{r}$. An active rod obeys Jeffery's equation \cite{jeffery1922motion,karrila1992microhydrodynamics} where the rod reorients due to background shear and random reorientation with diffusion coefficient $D_{\text{rot}}$. These equations hold when the active rod swims in the bulk, away from a microchannel's confinement. We model confinement by imposing that the active rod cannot penetrate the boundary, however, the rod can still swim along the wall and reorient. The non-dimensionalization of the model equations in Eq.~\eqref{eq:modelEquations} is shown in the Appendix. Hereafter, all simulation results are presented in their non-dimensionalized form with characteristic time and length scales of $2.5$ seconds and $50 \mu\mathrm{m}$, respectively, based on measurements of \textit{Bacillus subtilis} \cite{haines2009three,sokolov2009reduction,ryan2011viscosity}. 
    
    The background flow $\boldsymbol{u}$ is derived from the solution of the Stokes equation
    \begin{eqnarray}\label{eq:stokes}
        \mu \Delta \boldsymbol{u} + \nabla p = 0, & \nabla \cdot \boldsymbol{u} = 0
    \end{eqnarray}
    with the no slip boundary conditions on the confinement walls. For infinite cylindrical channels, to describe pressure-driven Poiseuille flow we represent the pressure gradient as $\nabla p = \alpha \boldsymbol{e}_z$ where $\boldsymbol{e}_z$ is the unit direction vector of the major axis of the microchannel and $\alpha$ is a prescribed parameter representing the pressure drop per unit length. This yields a solution to Eq.~\eqref{eq:stokes} of $\boldsymbol{u} = w(x,y)\boldsymbol{e}_z$ with $w(x,y)$ solving the two-dimensional Poisson's equation 
    \begin{eqnarray}\label{eq:poisson}
    \Delta w = \frac{\alpha}{\mu} \text{ in } \Omega, & w\vert_{\partial \Omega} = 0,
    \end{eqnarray}
    where the 2D domain $\Omega$ describes the cross-sectional geometry of the three-dimensional microchannel $M$ which then can be represented as $M=\Omega \times \{-\infty<z<\infty\}$. The total volume of fluid flowing through a cross-section per unit time is then 
    \begin{equation}\label{eq:totalFlow}
        \nu = \int\limits_\Omega |w(x,y)|\,\mathrm{d}x\mathrm{d}y.
    \end{equation}
    If $w_E(x,y)$ solves Eq.~\eqref{eq:poisson} where $\Omega$ is the ellipse with semi-axes $a$ and $b$, then the quantity in Eq.~\eqref{eq:totalFlow} can be expressed in terms of its centerline speed $|w_E(0,0)|$ as
    \begin{equation}\label{eq:flowEllipse}
        \nu_E = \frac{\pi ab}{2}|w_E(0,0)|.
    \end{equation}
    Similarly, for rectangular cross section, Eq.~\eqref{eq:totalFlow} can be obtained as 
    \begin{equation}\label{eq:flowRect}
    \nu_R = \kappa_R |w_R(0,0)|,
    \end{equation}
    where $|w_R(0,0)|$ is the respective centerline flow speed down the rectangular channel and $\kappa_R$ is written as follows \cite{tabeling2023introduction}
    \begin{equation}\label{eq:kappa_R}
    \kappa_R = \sum\limits_{n=1}^{\infty}\dfrac{4 b^2}{\pi (2n-1)^4}\left(\dfrac{2a}{b}-\dfrac{4}{(2n-1)\pi}\tanh\left(\dfrac{(2n-1)\pi}{2b}a\right)\right).
    \end{equation}
    

\begin{figure}[t!]
\centering
\includegraphics[width=1\linewidth]{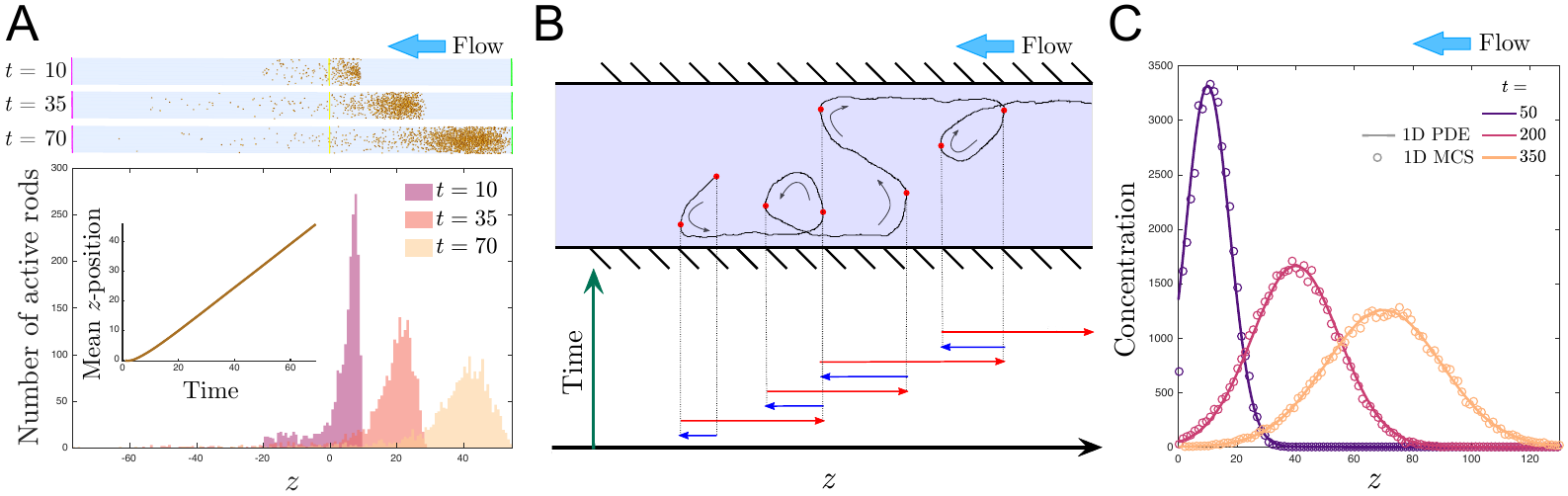}
\caption{{\bf Three models for transport of active rods swimming in microchannel.} \textbf{(A)} Snapshots of $R = 2000$ simulated active rods swimming inside the 3D elliptic microchannel (the 3D model from Subsection~\ref{subsec:3d-model}) shown at representative times $t = 10,35,$ and $70$. All $2000$ rods are initialized at $z = 0$ with random coordinates in the $xy-$plane and with random orientation. Each active rod is represented as half red and half green segments, where red depicts the front of the rod. The blue arrow at the top indicates the right-to-left direction of the background flow $\boldsymbol{u}$. Bottom plot shows histograms of rods' location along major axis $z$ at each representative time point. \textbf{(B)} Top: an example trajectory of an active rod swimming in the 3D elliptic microchannel in the $xz-$plane. Black arrows indicate the direction of the rod's movement. Bottom: projection of the rod's trajectory onto the $z-$axis only to illustrate 1D model for active rod transport (1D biased random walk from Subsection~\ref{sec:modelB}), with downstream and upstream displacements indicated by the blue and red arrows, respectively. \textbf{(C)} Representative simulation of the 1D models for active rod transport, comparing Monte Carlo simulations of 1D biased random walk model and the corresponding advection-diffusion equation \eqref{pdf-advection-diffusion}. Figure shows the simulated concentration profiles at three time points for $R = 50,000$ active rods initialized at $z = 0$. Circles correspond to Monte Carlo simulations (MCS) according to dynamics of the 1D biased random walk model. Here, the probability distributions for upstream and downstream persistence times were taken as a decaying exponential with mean $\tau_{\rm up} = 1.0$ and $\tau_{\rm down} = 0.5$, respectively. The thicker continuous line corresponds to the advection diffusion PDE \eqref{pdf-advection-diffusion} with $D_{\rm eff}$ and $V_{\rm eff}$ as in Eqs. \eqref{eq:effective_diffusion} and \eqref{eq:effective_velocity}, respectively. In all plots, concentration is the number of active rods counted per unit length.}
\label{fig:fig1}
\end{figure}

\subsection{A one-dimensional biased random walk model}
\label{sec:modelB}

Active rods in the 3D microfluidic channel typically follow the same general motility pattern: they swim upstream at the wall, escape the wall by random reorientation due to rotational diffusion, and then are transported downstream in the fluid bulk. After being advected downstream by the flow, the rod will again contact the wall and recommence its upstream motion. An example of such a trajectory is shown at the top of Figure~\ref{fig:fig1}(B), where the rod switches from swimming upstream close to the wall and then downstream in the fluid bulk. Furthermore, when reduced to the channel's major axis $z$ only, the trajectory of an active rod involves many successive directional reversals that characterize the rod's alternation between upstream swimming and downstream advection. This reduction of an active rod's trajectory to the $z$-axis is also illustrated in Figure~\ref{fig:fig1}(B). Thus, to study upstream transport dynamics of active rods swimming in the 3D microchannel, we consider the following simplified 1D biased random-walk model similar to the one proposed in \cite{figueroa2020coli}.

In this model, an active rod enters the channel and begins swimming upstream for time $t_{u}$ with constant speed $V_{u} > 0$. Then, after upstream swimming time expires, the rod turns around and moves downstream for time $t_{d}$ with speed $V_{d} > 0$ corresponding to its downstream transport by the fluid flow. After time $t_{d}$, the rod turns upstream and the process repeats. For each successive reversal, the rod's directional persistence time is independently sampled from a probability distribution function, i.e., $t_{u} \sim \psi_{u}(t)$ and $t_{d} \sim \psi_{d}(t)$. If the randomly generated upstream and downstream times are $t^{(1)}_{u},t^{(2)}_{u},...$ and $t^{(1)}_{d},t^{(2)}_{d},...$, then the $z$-location of the active rod after $N$ cycles of reversals is 
\begin{equation}
    z(T) = \sum\limits_{i=1}^N (V_{u}t^{(i)}_{u} - V_{d}t^{(i)}_{d}) \textrm{ where } T = \sum\limits_{i=1}^N (t^{(i)}_{u} + t^{(i)}_{d})
\end{equation}
Numerically, these times are sampled using an acceptance-rejection algorithm, and the simulated 1D trajectories of many active rods are stored and accumulated into a histogram to demonstrate the spatio-temporal transport dynamics. A representative concentration profile from the 1D Monte Carlo simulations is shown in Figure~\ref{fig:fig1}(C). In this simulation, the probability distributions of upstream and downstream swimming times were taken as decaying exponential functions: 
\begin{eqnarray}\label{exponential-distribution}
\psi_{u}(t) = \lambda_{u}\exp(-\lambda_{u} t), & \psi_{d}(t) = \lambda_{d}\exp(-\lambda_{d} t)
\end{eqnarray} 
In general, however, these times can be sampled from any continuous probability distribution function. Therefore, this reduced model provides a framework for capturing the core transport dynamics of active rods during upstream contamination while allowing us to examine the role of different persistence time distributions in governing population-level transport behavior. In Section~\ref{sec:3.2}, we address how the statistical properties of the probability distributions for rods' persistence times affect the qualitative shape of the moving concentration profiles. 

Such a generalization of random walk is also known as two-state continuous-time-random walk \cite{weiss1994aspects,metzler2000random,masoliver2017continuous} and velocity-jump process \cite{othmer1988models,othmer2000diffusion,othmer2002diffusion}. A key step in analyzing these processes is the Montroll--Weiss equality \cite{montroll1965random, weiss1994aspects}, which provides a closed-form expression for the joint Fourier--Laplace transform in space and time, respectively, of the probability distribution function $ p(z, t) $ of an individual walker. We show derivation of the Montroll--Weiss equality for the model described in this section in Appendix~\ref{appendix:velocity-jump}, see Eqn.~\eqref{general_FL_total_density}. 

For certain choices of the persistence-time distributions $ \psi_{u}(t) $ and $ \psi_{d}(t) $, taking the inverse transforms yields a PDE for $ p(z, t) $. For example, define $p_{u}(z,t)$ and $p_{d}(z,t)$ to be the probability densities of finding an active rod swimming upstream and downstream, respectively, at time $t$ and at location $z \in \mathbb{R}.$
Assuming exponential persistence-time distributions of the form \eqref{exponential-distribution} the evolution equations for $p_{u}$ and $p_{d}$ are shown in the Appendix~\ref{appendix:exponential} to be
\begin{equation}\label{eq:up_down_system}
    \begin{cases}
        \partial_t p_{u} + V_{u}\partial_z p_{u} = \lambda_{d}p_{d} - \lambda_{u}p_{u}, \\
        \partial_t p_{d} - V_{d}\partial_z p_{d} = \lambda_{u}p_{u} - \lambda_{d}p_{d},
    \end{cases}
\end{equation}
where the parameters $\lambda_{u}$ and $\lambda_{d}$ are the rates at which a rod reverses its direction. In particular, if $\tau_{u}$ and $\tau_{d}$ are the mean times that rods spend swimming upstream and downstream, then $\lambda_{u} = \tau_{u}^{-1}$ and $\lambda_{d} = \tau_{d}^{-1}$. Furthermore, the total probability density for a rod's location $z$ after time $t$ is then $p(z,t) \equiv p_{u} + p_{d}$ and, when $V_{u} = V_{d} = V$, the system Eq. \eqref{eq:up_down_system} can be written as a single PDE for $p$ as follows: 
\begin{equation}\label{eq:telegraphers}
   \partial^2_t p + (\lambda_{u} + \lambda_{d})\partial_t p = V^2\partial^2_z p - V(\lambda_{d} - \lambda_{u})\partial_z p.
\end{equation}
This equation, equivalent to the system \eqref{eq:up_down_system}, is known as the (biased) telegrapher's equation and has arose in previous studies of persistent one-dimensional motion, such as the run-and-tumble motility of bacteria and bacterial chemotaxis \cite{othmer1988models, rivero1989transport, ahmed2008experimental, othmer2000diffusion, erban2004individual, schnitzer1993theory, tailleur2008statistical}. Furthermore, when timescales much longer than the mean reversal time are considered ($t \gg (\lambda_{u} + \lambda_{d})^{-1}$), the telegraphers equation Eq. $\eqref{eq:telegraphers}$
reduces to the standard advection-diffusion equation (see also Appendix~\ref{appendix:exponential})
\begin{equation}
\label{pdf-advection-diffusion}
    \partial_tp + V_{\rm eff}\partial_zp = D_{\rm eff}\partial_z^2 p,
\end{equation}
where the effective coefficients $D_{\rm eff}$ and $V_{\rm eff}$ are as follows: 
\begin{equation} \label{eq:effective_diffusion}
    D_{\rm eff} = \frac{V^2}{\lambda_{u} + \lambda_{d}} = V^2 \dfrac{\tau_{u}\tau_{d}}{\tau_{u} + \tau_{d}}
\end{equation}
and
\begin{equation} \label{eq:effective_velocity}
    V_{\rm eff} = V\frac{\lambda_{d} - \lambda_{u}}{\lambda_{d} + \lambda_{u}} = V\dfrac{\tau_{u} -  \tau_{d}}{\tau_{u} + \tau_{d}}.
\end{equation}
The value of $V_{\text{eff}}$ represents the average velocity. Indeed, over two consecutive runs—one downstream and one upstream—the active rod experiences an average displacement of $V_{u}\tau_{u} - V_{d}\tau_{d} = V(\tau_{u} - \tau_{d})$, and the total time elapsed is $\tau_{u} + \tau_{d}$. We note that an alternative coarse-graining approach for active Brownian particles, based on a small-wavenumber expansion of the Smoluchowski equation, has been developed to characterize effective advection-diffusion transport of these particles in Poiseuille flow \cite{peng2020upstream}.

Assuming the active rod initiates its motion at $z=0$, we impose the initial conditions $p|_{t=0}=\delta(z)$. This initial-value problem admits the well-known analytical solution for $t>0$: 
\begin{equation}
\label{analytical-solution-of-advection-diffusion-equation}
    p(z,t) = \dfrac{1}{\sqrt{4\pi D_\text{eff}t}}\text{exp}\left\{-\dfrac{(z-V_{\text{eff}}t)^2}{4D_{\text{eff}}t}\right\}.
\end{equation}

In Appendix~\ref{appendix:exponential} we present the equation for $p$ with its derivation for the case when downstream and upstream speeds do not necessarily coincide. For more general time distributions $\psi_{d}(t)$ and $\psi_{u}(t)$, even those with exponential tails, the resulting equations may become significantly more complex. For instance, when $\psi_{d}(t) = \psi_{u}(t) = \lambda^2 t e^{-\lambda t}$, the governing equation becomes third-order in time and fourth-order in space \cite[Eqn.~(48)]{masoliver2017continuous}. In the case of polynomial-tail (or ``heavy-tail") distributions $\psi_{d}(t)$ and $\psi_{u}(t)$, the resulting equation involves a fractional time derivative, \cite[Section 3.4]{metzler2000random} and \cite[Section 5]{masoliver2017continuous}. 
Finally, in Appendix~\ref{appendix:deterministic} we derive a partial differential equation for the probability distribution function $p(z,t)$ in the case where the downstream swimming time is deterministic, i.e., $\psi_{d} (t)= \delta(t - T_d)$; see Eq.~\eqref{eqn_for_p_in_deterministic_case}. Besides the presence of memory terms in the corresponding equation, this specific choice of waiting-time distribution is of interest because it was used in \cite{figueroa2020coli} to compare simulations of the biased random-walk model with experimentally derived bacterial profiles.

\begin{figure}[t!]
    \centering
    \includegraphics[width=1\textwidth]{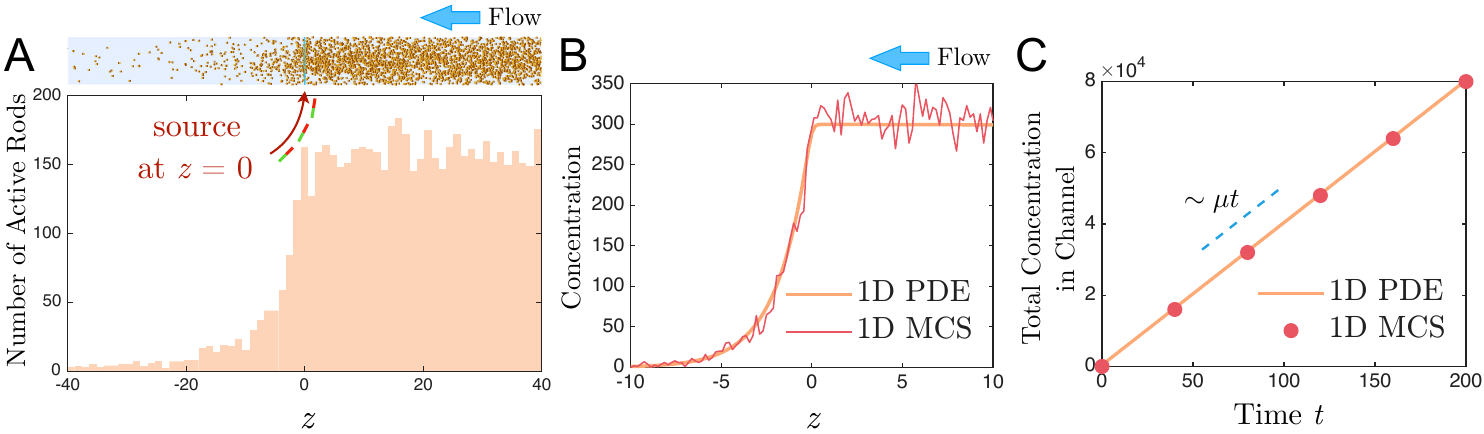}
    \caption{\textbf{Illustration of the contamination setup.} \textbf{(A)} Numerical simulation of the 3D model of active rods in a cylindrical microchannel under the constant influx of active rods at the point $z = 0$, reflecting the experimental setup from \cite{figueroa2020coli}. The resulting concentration profile along the major axis $z$ is shown near the site of injection. \textbf{(B)} Comparisons of concentration profiles for biased random walk model (1D MCS, where MCS stands for Monte Carlo simulations) and advection-diffusion equation \eqref{pdf-advection-diffusion} (1D PDE) near $z=0$ with injection frequency $\mu$. Effective diffusivity and velocity for the PDE model (Eq.~\eqref{advection-diffusion-equation-for-c}) for concentration $c$ are taken as in Equations~\eqref{eq:effective_diffusion} and \eqref{eq:effective_velocity}, respectively. \textbf{(C)} Total concentration of active rods injected as a function of time. For the PDE model, this is computed as $\int_\mathbb{R} c(x,t)\,\mathrm{d}x$ which is seen to increase linearly with injection frequency $\mu$.}
    \label{fig:contamination}
\end{figure}

In microfluidic experiments of bacterial contamination \cite{figueroa2020coli}, a reservoir of bacteria at the fluid outlet resulted in an influx of new swimmers into the channel over time. To model this experimental setup, rods are initiated at $z=0$ every $\tau_0$ units of time. An illustration of this setup, together with the resulting concentration profiles near the injection site, is shown in Figure~\ref{fig:contamination}. In this case, we can then define the concentration profile of active rods as
\begin{equation*}
c(z,t) = \mathbb E \left[\sum\limits_{m=1}^{N_t}\delta(z-z_m(t))\right]=\sum\limits_{m=1}^{N_t}p(z,t-m\tau_0).
\end{equation*}
Here, $N_t=\left\lfloor \frac{t}{\tau_0} \right\rfloor$ is the number of active rods initiated through time interval $(0,t)$ and $z_m(t)$ is the coordinate of active rod's trajectory along the microchannel's major axis. Note that the function $z_m(t)$ is a random variable for all $t>m\tau_0$ and its probability distribution function is $p(z,t-m\tau_0)$. If we assume that active rods are injected continuously at the rate $\mu = \tau^{-1}_0$, then 
\begin{equation}\label{c_via_p}
c(z,t) = \mu \int\limits_0^t p(z,t-\tau)\,\text{d}\tau=\mu\int\limits_0^tp(z,\tau)\,\text{d}\tau.
\end{equation}
This formula relates the concentrations of active rods in two different scenarios: when all rods are initially located at $z=0$ and when rods are injected at $z=0$ at a frequency $\mu$.

In the case when $p$ satisfies the advection-diffusion equation \eqref{pdf-advection-diffusion}, using integration in time of the both sides of equation \eqref{pdf-advection-diffusion}, $\partial_t  c = \mu p $, and $\int\limits_0^t \partial_t p(z,\tilde{t}) \,\text{d}\tilde{t}= \mu^{-1}\partial_t c -\delta(t)$, one can show that the concentration profile $c(z,t)$ solves 
\begin{equation}\label{advection-diffusion-equation-for-c}
\partial_t c + V_{\mathrm{eff}}\partial_z c = D_{\mathrm{eff}}\partial_z^2 c+\mu \delta(z),
\end{equation}
subject to the zero initial condition $c|_{t=0}=0$. This initial-value problem also admits the analytical solution 
\begin{equation}\label{formula-for-c}
c(z,t) = \int\limits_0^t \dfrac{\mu}{\sqrt{4\pi D_{\text{eff}}\tau}}\exp\left\{-\dfrac{(z-V_{\text{eff}}\tau)^2}{4D_{\text{eff}}\tau}\right\}\,\text{d}\tau,
\end{equation}
which one can derive by simply using equality \eqref{c_via_p} and the analytical solution \eqref{analytical-solution-of-advection-diffusion-equation} for~$p(z,t)$.

\section{Results}
\subsection{Dependence of contamination on flow rate, rotational diffusion, and the microchannel’s cross-sectional geometry.}\label{sec:3.1}

Simulation results from the 3D model of active rods in a cylindrical microchannel clearly demonstrate the net upstream progression of active rods in the microchannel due to negative rheotaxis (see Figure~\ref{fig:fig1}(A)). Additionally, the spatio-temporal profiles show a well-defined front where the active rod population is most dense that propagates upstream, followed by a fast decline of rod density in regions of the channel further downstream. Characteristic of an advection-diffusion process, the rod concentration profile along the major axis $z$ widens and drifts to the right, against the background flow $\boldsymbol{u}$. In this case, when the active rod population exhibits net migration against the background flow and gradually invades upstream regions over time, we refer to it as \textit{contamination} as in \cite{figueroa2020coli}. Furthermore, as shown in the inset of Figure~\ref{fig:fig1}(A), after a short transient time, the active rod population's mean location along the channel increases linearly with time, suggesting that the  rods' upstream transport is characterized by an effective velocity, $V_{\rm eff}$. In this section, we investigate how this effective velocity as well as rods' directional persistence times depend on the flow strength, rotational diffusion, and the microchannel's cross-sectional geometry. 

\begin{figure}
    \centering
    \includegraphics[width=0.90\textwidth]{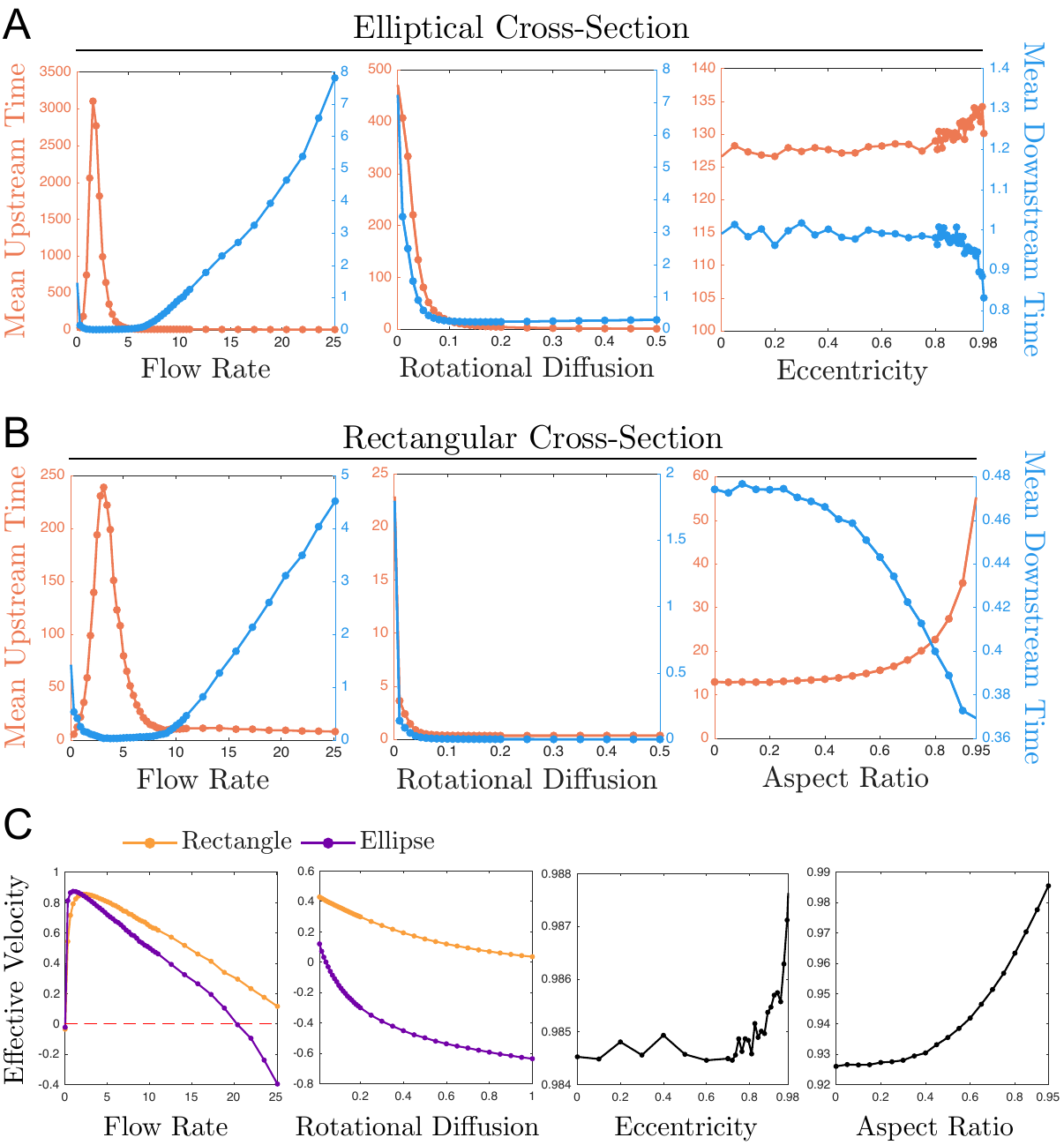}
    \caption{{\bf Dependence of active rods' directional persistence times and effective velocity on flow rate, rotational diffusion, and aspect ratio in microchannels of elliptical and rectangular cross-section.} In A (elliptical channel) and B (rectangular channel), the mean upstream time (left/orange axis) and mean downstream time (right/blue axis) are computed as the average of all upstream and downstream persistence time intervals before direction reversal over $R = 20,000$ active rods. Notice the difference in magnitude between the scales of the upstream (left) and downstream (right) axes. In left panels, the flow rate value being varied is centerline speed, $|w_E(0,0)|$ and $|w_R(0,0)|$, but the axes are rescaled to show the total volume of fluid passing through the cross-section per unit time. Effective velocity in subfigure (C) is computed as a rod's final position minus its initial position divided by total simulation time $T$, averaged across $R=20,000$ active rods. When one parameter is varied, all others are kept at a fixed typical value, which we take as: $|w_E(0,0)| = 0.5, |w_R(0,0)| = 0.3608, D_{\rm rot} = 0.035, e=0.97,$ and $\alpha = 0.25$. When eccentricity and aspect ratio are varied, the area parameter $s=ab$ is fixed as $s=4$ so that the total flow rate remains the same. Aspect ratio is taken to be $1-\alpha$, so that the channel is most narrow at aspect ratio $\sim 1$ and square for aspect ratio $0$.}
    \label{fig:parameters}
\end{figure}

To quantify  the rods' net displacement through the channel, we consider the effective velocity, $V_{\rm eff}$, defined as each rod's final position minus initial position over total time, averaged across all active rods. When $V_{\rm eff} > 0$, rods on average are displaced upstream against the current indicating that the contamination occurs. If $V_{\rm eff} < 0$, no contamination occurs. Additionally, to study statistics of active rods' trajectories, we compute their persistence time intervals before full reorientation in the $z$ direction. In other words, we record the time that each rod swims upstream and downstream before turning around -- i.e., while $z(t) > z(t-\delta t)$ and $z(t) < z(t-\delta t)$. We then compute the averaged upstream and downstream persistence times across all rods, denoted $\tau_{\rm up}$ and $\tau_{\rm down}$, respectively. As shown in the next section, these quantities provide an effective parameterization of both the biased random walk model and the corresponding PDE for the probability density $p(x,t)$, allowing us to approximate the spatio-temporal profiles of upstream contamination.


We first investigate how the flow rate of the background fluid affects active rods' upstream and downstream persistence times in microchannels of both elliptical and rectangular cross-section. To make an accurate comparison between the two channels, we must first ensure that the volume of fluid passing through the ellipse and rectangle per unit of time is kept the same. This amounts to equating Eq.~\ref{eq:flowEllipse} with Eq.~\ref{eq:flowRect} so that, as we vary the flow speed down the centerline in the elliptical channel, the corresponding centerline speed in the rectangular channel is 
\[|w_R(0,0)| = \frac{\pi ab}{2\kappa_R}|w_E(0,0)|.\]
When there is zero background flow, rods will not exhibit the negative rheotaxis nor the downstream advection that drives the transport of active rods up and down the channel. As a result, we observe average upstream and downstream persistence times of $\sim 1$. In this case, since rods will on average swim upstream for as long as they do downstream, the resulting effective velocity is $V_{\rm eff} \approx 0$, indicating no net upstream progression and thus no contamination. As the background flow speed increases, the mean upstream times also increase to a maximum, peak value. For the elliptical microchannel, this peak in persistence occurs around centerline flow rate of $|w_E(0,0)|=0.25$, or a total flow rate of $\nu_E \approx 1.5708$ (Figure~\ref{fig:parameters}(A), left). In rectangular channel, the peak occurs at an approximate flow speed of $|w_R(0,0)|=0.36$, corresponding to $\nu_R = 3.1415$ (Figure~\ref{fig:parameters}(B), left). When translated back into its dimensional form, these peaks where rods are the most persistent upstream swimmers occur at flow rates of $1570.8 \mu\rm{m}^2/s$ in the elliptical channel, and $ 3141.5\mu\rm{m}^2/s$ in the rectangular channel. As flow rate is increased further, rods in both channels move upstream with less persistence and spend more time being advected downstream by the flow. At low flow rates, upstream persistence times in the elliptical channel are much higher than those in the rectangular channel. Moreover, the effective velocity in the channel of rectangular cross-section is initially smaller when the fluid flow rate is low, indicating that rods contaminate the channel slower than they do in channels of elliptical cross-section (Figure~\ref{fig:parameters}(C), left). However, as flow rate increases, rods' effective velocity in rectangular channels quickly exceeds effective velocity in the elliptical channel. In both geometries, we observe a net-progress upstream ($V_{\rm eff} > 0$) for a large range of flow rates. The effective velocity remains higher in the rectangular channel at large flow rates and $V_{\rm eff}$ doesn't become negative until a flow rate of $>25$, whereas in the elliptical channel, we observe $V_{\rm eff} < 0$ after a flow rate of around $21$. This faster contamination rate of rectangular channels is further exemplified later by a comparison of simulated concentration profiles between the two channels ran at a higher flow rate (Figure~\ref{fig:plateauProfiles}(C)-(D)). In the rectangular channel, we observe rods at the front of the population advancing approximately $25\%$ further than in the elliptical channel after the same, short amount of time. 


Next, we study how changes to a rod's rotational diffusion can affect its directional persistence. When $0 < D_{\rm rot} \ll 1$, an active rod that makes contact with the channel wall will tend to remain there and continue swimming in its current upstream direction for longer. This is because fluid flow and wall interactions alone are insufficient to reorient the rod; without rotational diffusion, the rod has no mechanism to escape the wall and reorient towards the fluid bulk to reverse its direction. Similarly, rods that initially swim downstream will take much longer to reorient and swim upstream. Thus, for very small $D_{\rm rot}$, we observe highly persistent rods that exhibit large upstream and downstream turnaround times (Figure~\ref{fig:parameters}(A-B), center panels). As $D_{\rm rot}$ increases, however, reorientation become more frequent and the mean time traveling before a rod reverses its direction quickly decreases. Moreover, the decreased directional persistence by increasing $D_{\rm rot}$ results in lower effective velocity of active rods through the channel (Figure~\ref{fig:parameters}(C)). While effective velocity is initially higher in the elliptical channel for small rotational diffusion, this decrease of effective velocity happens slower in the rectangular channel as $D_{\rm rot}$ increases. 

We also examine the effects of the microchannel's cross-sectional aspect ratio on active rods' ability to persist upstream against the flow. For a channel of elliptical cross-section, this amounts to varying the ellipse's eccentricity $e = \sqrt{a^2-b^2}/a$. We fix the quantity $s\coloneqq ab$ so that the cross-sectional area of the ellipse remains the same as $e$ is varied. Since $\nu_E$ is directly proportional to the product $ab$ (Eq.~\eqref{eq:flowEllipse}), fixing $s$ ensures that the total amount of fluid flowing through the channel's cross section is the same across all eccentricity values. Moreover, for fixed area $s=ab$ and eccentricity $e$, we have that
\[a = \sqrt{s} (1-e^2)^{-1/4} \text{ and } b = \sqrt{s} (1-e^2)^{1/4}.\]
In the case of a rectangular cross-section, we similarly define a parameter $\alpha = a/b$ and keep the area parameter $s=ab$ fixed. In this case, we can obtain expressions for $a$ and $b$ as 
\[a = \sqrt{s \cdot \alpha} \text{ and } b = \sqrt{\frac{s}{\alpha}}.\]
Then, $\alpha = 1$ corresponds to square channels whereas $0<\alpha\ll 1$ corresponds to highly rectangular channels. However, for the sake of comparison with elliptical channels, we define the aspect ratio as $1-\alpha$ so that aspect ratio of $0$ ($\alpha = 1$) corresponds to square channels and aspect ratio $\approx 1$ ($\alpha \approx 0$) corresponds to highly rectangular channels. Under this convention, the eccentricity/aspect ratio values range from $0$ (circle/square) and increase toward $1$ as the cross-sections become increasingly elongated. 

As can be seen from Eq.~\eqref{eq:kappa_R}, the fluid flow rate in the rectangular channel does not scale linearly with the area $s = ab$ like it does in the elliptical channel. To ensure the volume of fluid passing through the channel, $\nu_R$, remains the same for all $\alpha$, we must also solve for the corresponding centerline speed $|w_R(0,0)|$ as 
\[|w_R(0,0)| = \frac{\nu_R}{\kappa_R}.\] 
In our simulations, we fix $s = 4$ and we vary $e \in [0,0.98]$, $\alpha \in [0.05,1]$. As seen in Figure~\ref{fig:parameters}(A), in the elliptical microchannel, the rods' mean upstream and downstream persistence times remain relatively constant across many values of the eccentricity ($0 \leq e \lessapprox 0.8$). Upstream persistence then increases as the microchannel becomes narrow for $0.8 \leq e \leq 0.98$, meanwhile the downstream persistence times decrease slightly. Consistent with mean upstream persistence times increasing and mean downstream persistence times decreasing, we also observe an increase in the effective velocity as the eccentricity parameter approaches $1$ (Figure~\ref{fig:parameters}(C)). 

In the rectangular microchannel, increasing aspect ratio $1-\alpha$ steadily increases the rods' upstream times, and decreases their downstream times (see Figure~\ref{fig:parameters}(B), right panel). Moreover, effective velocity increases as aspect ratio increases, implying that active rods' net upstream progress is greatest for highly rectangular cross-sections ($0 < \alpha \ll 1$), and least for microchannels with square cross-section ($\alpha = 1$). 

\begin{figure}[t!]
    \centering
    \includegraphics[width=1\textwidth]{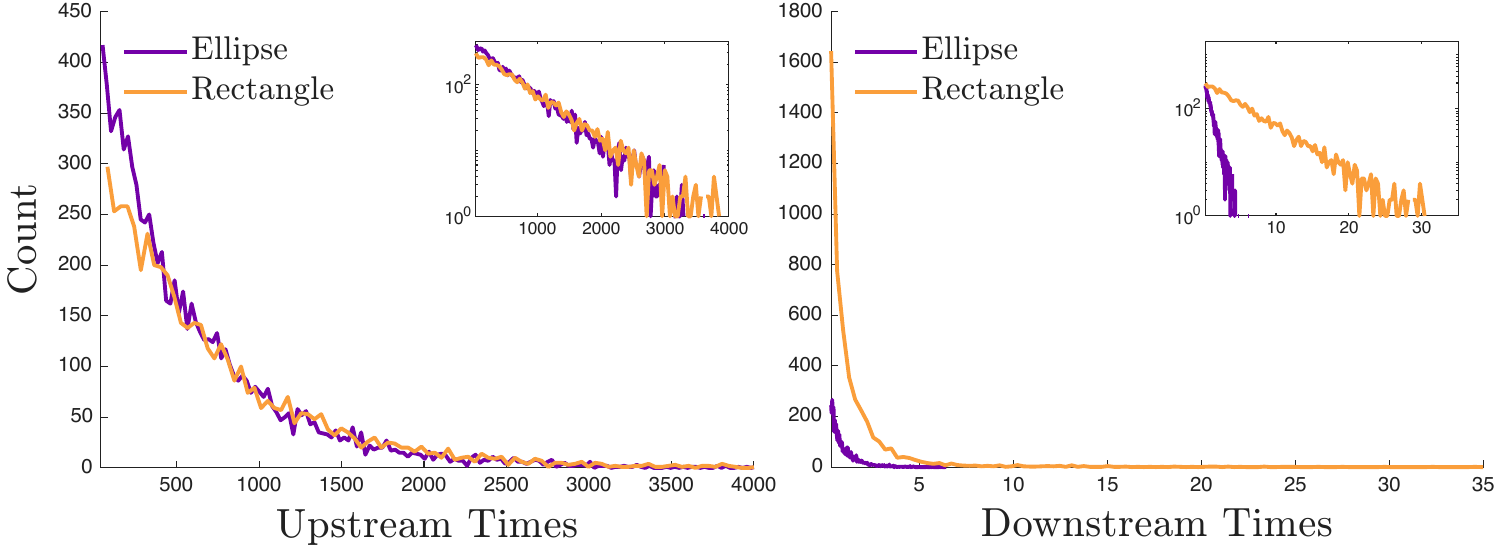}
    \caption{{\bf Distributions of active rods' directional persistence times in microchannels of elliptic and rectangular cross-section.} Simulations are ran with $R=20$ active rods until each rod has changed its $z$-direction $3,500$ times, producing $70,000$ total turnarounds and $35,000$ upstream and downstream persistence time intervals. Figure shows resulting histograms of directional persistence times as line plots with $\lceil \sqrt{35,000}\rceil$ bins. Parameter values are chosen as $D_{\rm rot} = 0.035, e = 0.97, \alpha = 0.25$, and $|w_E(0,0)| = 0.5$ corresponding to $|w_R(0,0)| = 0.3608$ so that the total flow rate is the same between the elliptic and rectangular microchannels. Insets show the same distributions as the main figure but on a logarithmic scaled $y-$axis.} 
    \label{fig:histograms}
\end{figure}

\begin{figure}[t!]
\includegraphics[width=0.725\textwidth]{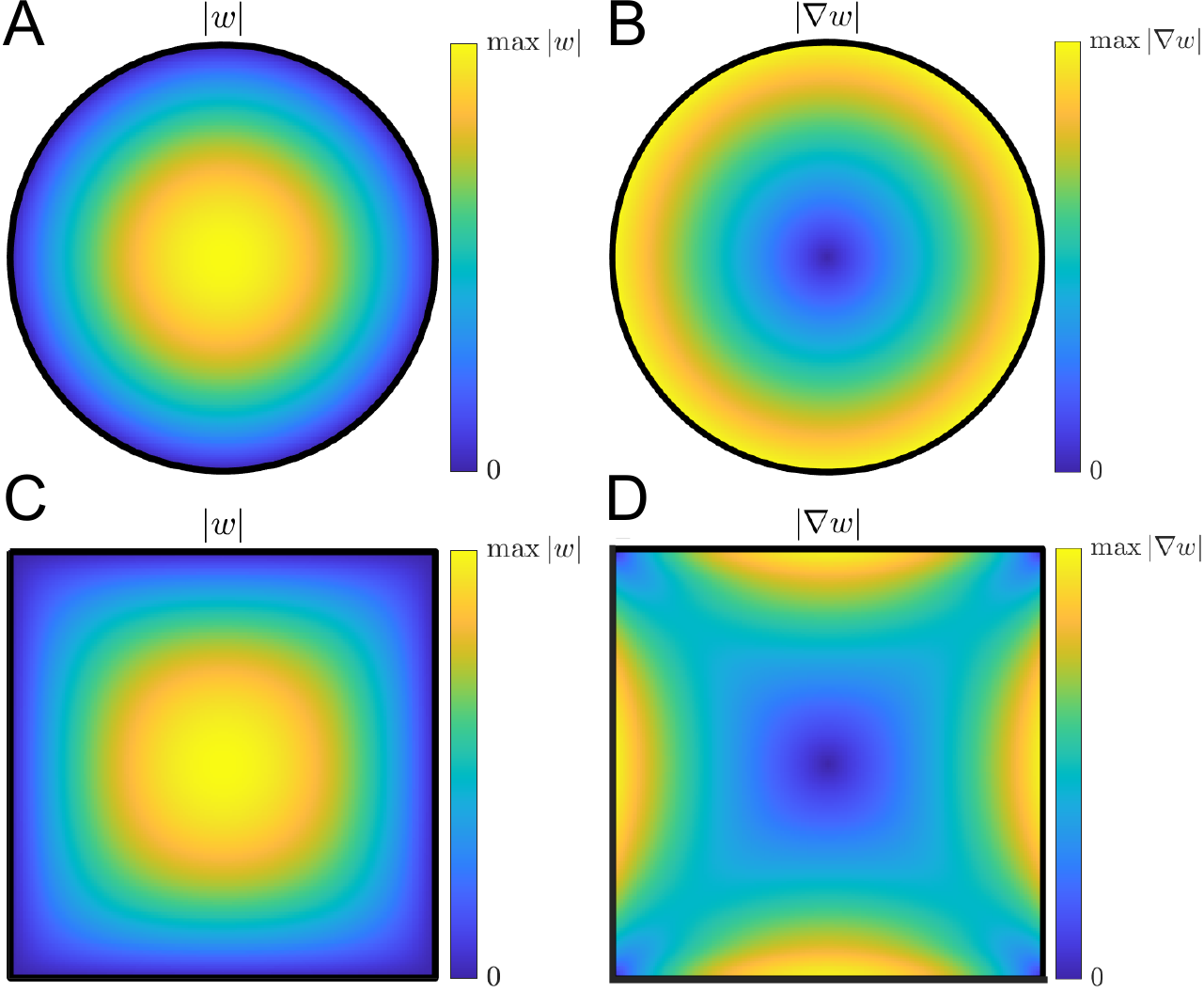}
\caption{{\bf Poiseuille flow in microchannels with circular and rectangular cross-sections: speed $|w|$ and local shear rate $|\nabla w|$}. In each sub-figure, a cross-section of a microchannel is depicted. Color represents the magnitude of the flow (A,C) and the local shear rate (B,D).   }
\label{fig:poiseuille}
\end{figure}

Finally, we utilize our 3D model of active rods in a cylindrical microchannel to investigate the distribution of active rods' upstream and downstream persistence times. Similar to the statistic for rods' directional persistence defined earlier, we recorded each time interval that rods swam before reversing their direction. The resulting histograms, shown in Figure~\ref{fig:histograms}, indicate that upstream and downstream persistence times of active rods follow a decaying exponential distribution. Furthermore, the corners of a rectangular microchannel are known to generate regions of low shear (see Fig.~\ref{fig:poiseuille}), where active rods tend to accumulate and swim along the walls. This situation contrasts with circular microchannels, in which the shear rate is highest and uniform along the entire boundary (one can compare shear rates for circular and rectangular microchannels in Figure~\ref{fig:poiseuille} B and D). The combination of reduced shear and corner-guided swimming in rectangular geometries may be conducive to highly persistent upstream swimming, motivating the potential use of a heavy-tailed power-law distribution to describe persistence time durations. Here, however, while we do observe instances of very long upstream swimming intervals, the overall distribution of persistence times remains exponential in microchannels of both elliptical and rectangular cross-section.

\begin{remark}
We conclude this section by explaining why persistence-time distributions decay exponentially. 
We focus on persistence times associated with upstream swimming, which predominantly occur when the active rod is located at the wall and oriented toward it (cf.\ the illustration in Fig.~\ref{fig:fig1}B). 
Accordingly, we approximate the upstream persistence time by the time required for the active rod to escape from the wall. 
The argument for the exponential decay of the escape-time distribution is based on its interpretation in terms of a Fokker--Planck equation with absorbing boundary conditions. 
Solutions of such problems are well known to exhibit exponential decay at long times \cite{risken1989fokker}.

For the sake of clarity, consider the two-dimensional active rod model, see, {\it e.g.}, \cite{potomkin2017focusing,rubio2021self}. If the active rod has collided with wall $\left\{x=0\right\}$ from above, subject to background shear flow $\boldsymbol{u}=(0,\,\gamma y)$, its orientational angle $\varphi$, while the rod is reorienting at the wall, is downward, that is, $\varphi \in (-\pi, 0)$, solves the following problem 
\begin{equation}\label{phi_stochastic_equation}
    \begin{cases}
        \mathrm{d}\varphi(t) = -\gamma\sin^2(\varphi(t))\mathrm{d} t +\sqrt{2D_{\mathrm{rot}}}\,\mathrm{d} W,\\
        \varphi(0)\sim\mathcal{U}(-\pi,0).
    \end{cases}
\end{equation}
Here, we assume that the active rod lands on the wall with orientation $\varphi(0)$ which is a random variable uniformly distributed on $(-\pi,0)$. The active rod is capable of escaping from the wall when $\varphi (t)$ goes outside the interval $(-\pi,0)$, so the active rod reorients away from the wall. We aim to estimate the time $T_{\text{esc}}$ active rod needs to escape from the wall. The time $T_{\text{esc}}$ is a random variable; denote its probability distribution function by $f(T)$.   
Consider the corresponding Fokker-Planck problem for probability distribution function $P(\varphi,t)$ for the orientational angle $\varphi$: 
\begin{equation}\label{phi_fokker_planck}
    \begin{cases}
        \partial_tP +\partial_{\varphi}\left(-\gamma\sin^2(\varphi)P\right)=D_{\mathrm{rot}}\partial_{\varphi}^2P,\\
        P|_{\varphi=-\pi}=P|_{\varphi=0}=0,\\
        P|_{t=0} \equiv 1/\pi. 
    \end{cases}
\end{equation}
The absorbing boundary conditions mean that if the active rod reaches the orientation $\varphi=0$ or $\varphi=
\pi$ it escapes and is no longer considered. In other words, the probability that the active rod has not escaped is 
\begin{equation}
\mathbb{P}\left\{t\leq T_{\text{esc}}\right\} =\int\limits_{t}^{\infty} f(T) \, \text{d}T = \int\limits_{-\pi}^{0}P(\varphi,t)\,\text{d}\varphi.
\end{equation}
By taking derivatives in $t$ we obtain the formula for probability distribution function:
$
    f(t) = -\int\limits_{-\pi}^0\partial_tP\, \text{d}\varphi.
$
Further, the solution to the Fokker-Planck problem \eqref{phi_fokker_planck} can be written as an eigenfunction expansion
$P(\varphi,t) = \Sigma_{k=1}^{\infty}e^{-\lambda_k t}\Phi_k(\varphi)\sim e^{-\lambda_1t}\Phi_1(\varphi)$ for large $t$, where $P$'s behavior over time will be dominated by it's first eigenfunction in time. Thus,
$    f(t)\sim e^{-\lambda_1t}$ for large~$t$, that is the probability distribution function of escape times $f(t)$ is asymptotically exponential for large escape times $t$.
\end{remark}

\subsection{Concentration profiles: Plateau vs Slope}\label{sec:3.2}

We now employ the 1D models outlined in Section \ref{sec:2} to study how the qualitative behavior of upstream concentration profiles depends on the statistical properties of the underlying probability distributions for rods' persistence times. In \cite{figueroa2020coli}, a similar model for the 1D upstream-downstream trajectories was formulated alongside experiments for \textit{Escherichia coli} bacteria swimming in the microfluidic channel. Experimentally, bacteria were steadily injected into the microchannel at the fluid outlet, $z=0$. From there, the bacteria began their upstream-downstream swimming trajectories and exhibited a net upstream displacement towards the fluid source. In this context, upstream swimming at the wall corresponded to the highly persistent ``runs'' of bacteria. On the other hand, reorientation and detachment from the wall occurred due to tumbling events, at which point bacteria were advected downstream by the flow. Moreover, the qualitative behavior of the bacterial population's upstream contamination was thought to depend critically on the run-and-tumble statistics of bacteria. To illustrate this, two different probability distributions for bacterial run-times were considered. The first distribution was a decaying exponential of the form
\begin{equation}\label{eq:exp-dist}
    \psi(t) = \frac{\exp\left( -\frac{t}{\tau} \right)}{\tau}
\end{equation}
where $\tau > 0$. This choice reflects the classical assumption that tumbling events follow a Poisson process of rate $\frac{1}{\tau}$ and thus that the intervals between two successive tumbles (i.e., the bacteria's run times) are distributed exponentially. The second probability distribution for run-times of bacteria was a power-law of the form
\begin{equation}\label{eq:PL-dist}
    \psi(t) = \frac{\gamma}{\tau\left( 1 + \frac{t}{\tau}\right)^{\gamma + 1}}
\end{equation}
where $\tau>0$ and $\gamma > 1$ are parameters of the distribution. Such a heavy-tailed distribution was experimentally found in \cite{korobkova2004molecular} to better approximate run-time distribution of \textit{E. coli} with $\gamma = 1.2$ and $\tau = 1$s. Compared to the exponential case Eq.~\eqref{eq:exp-dist}, the power-law distribution Eq.~\eqref{eq:PL-dist} is heavy-tailed and decays at a slower rate, allowing for a higher probability of observing longer run-times. The resulting simulated 1D concentration profiles in \cite{figueroa2020coli} corresponding to the two probability distributions for swimmer run-times displayed remarkably different qualitative behavior. For exponentially distributed run times, Eq.~\eqref{eq:exp-dist}, the concentration of bacteria plateaued, remaining relatively constant over the channel's $z$-axis before steeply dropping off at the front. With the power-law distribution, the concentration profile did not plateau but instead consisted of slopes exhibiting slow monotonic decay of the bacterial concentration as distance from the bacterial reservoir increased. The simulated concentration profiles corresponding to power-law distributed run-times displayed stronger agreement with experimental concentration profiles. 

To further investigate the role of distributions for swimmers' persistence times on the contamination process at the population level, we first employ the 1D biased random walk model for active rod transport under the continuous injection of new swimmers into the microchannel (outlined in Section~\ref{sec:modelB} and shown in Figure~\ref{fig:contamination}). We also simulated 1D PDE \eqref{advection-diffusion-equation-for-c} for concentration $c$. We let the singular term in Eq.~\eqref{advection-diffusion-equation-for-c} take the form of an approximation of the delta function:

\[\delta(z) \approx \mu\frac{1}{\sqrt{2\pi}\varepsilon} e^{\frac{-z^2}{2\varepsilon^2}},\] for small $\varepsilon > 0$. 

To simulate rods' contamination when the underlying time distribution obeys a power-law, we use measurements of bacterial run-times from \cite{korobkova2004molecular}, which reported a distribution as in Eq.~\eqref{eq:PL-dist} with $\gamma = 1.2$ and $\tau = 1$s. This results in a mean upstream swimming time of $\tau_{\rm up} = \frac{\tau}{\gamma-1} = 5$s and divergent second moment. For consistency, in the exponential case, we also use $\tau_{\rm up} = 5$s which has a finite second moment of $50\rm{s}^2$. As shown earlier in Figure~\ref{fig:parameters}(A)-(B), at low flow rates, downstream persistence times are small and several times less than upstream swimming times. We thus choose an average downstream swimming time of $0.5$s, corresponding to $\tau_{\rm down} = 0.5$s in the exponential case and $(\gamma,\tau) = (3,1\rm{s})$ for the power-law case. 

\begin{figure}[t!]
    \centering
    \includegraphics[width=1\textwidth]{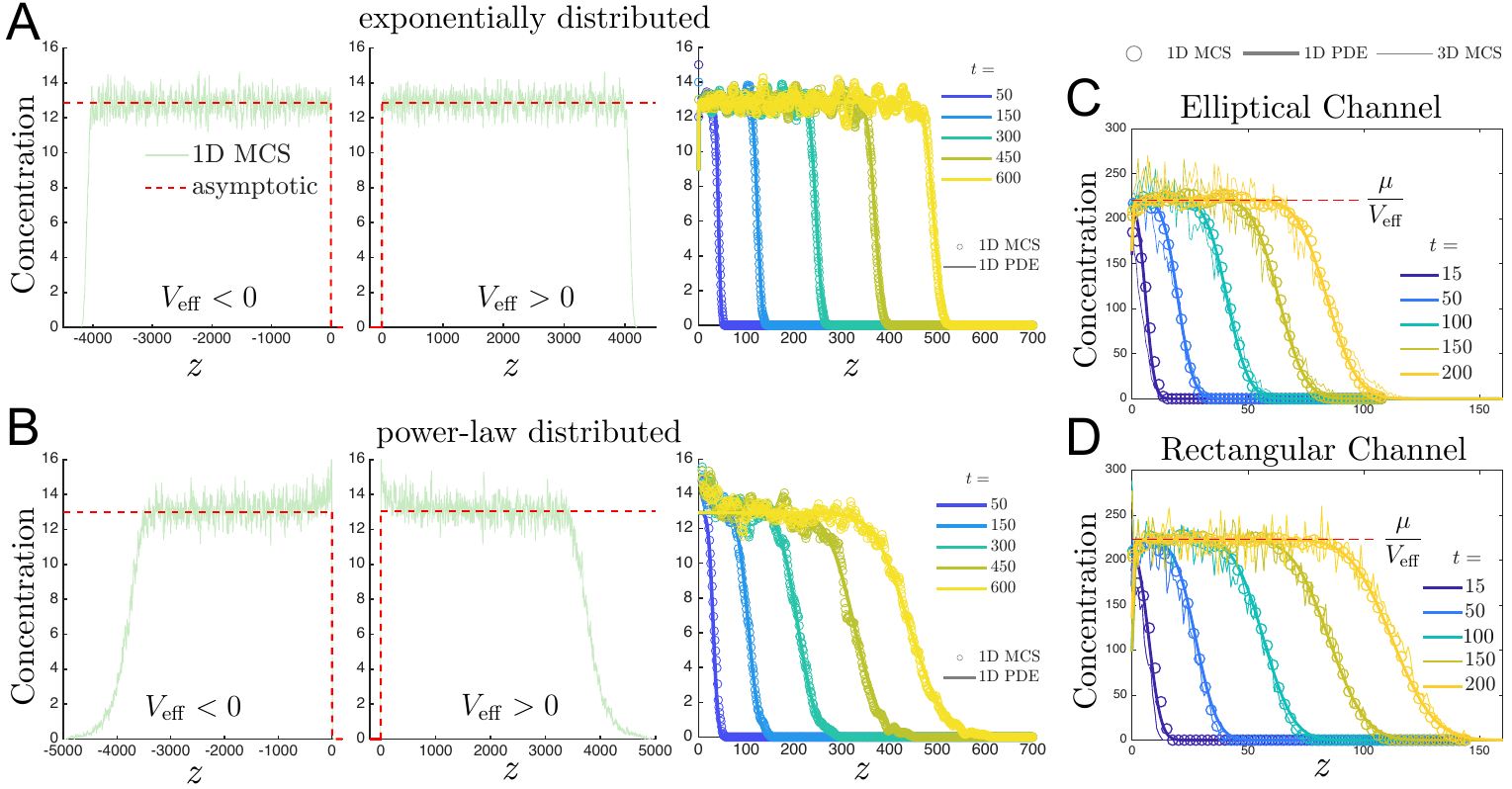}
    \caption{{\bf 1D concentration profiles of the contamination process.} \textbf{(A)}-\textbf{(B)} Comparison of concentration profiles from 1D Monte Carlo simulations of the biased random walk model (abbreviated 1D MCS) with the expression derived in Eq.~\eqref{eq:profileShape} (red dashed line). The right panel of (A) and (B) compares the spatiotemporal concentration profiles from the 1D biased random walk (circles) and the advection-diffusion PDE (thick continuous line). Simulations are performed for $R=150,000$ active rods. In (A), distributions for both upstream and downstream persistence times are exponential as shown in Eq. \eqref{eq:exp-dist} with $\tau_{\rm up} = 5\rm{s}$ and $\tau_{\rm down} = 0.5\rm{s}$ for $V_{\mathrm{eff}} > 0$, and the values are switched for $V_{\rm eff} < 0$. In (B), distributions follow a power-law as in Eq.~\eqref{eq:PL-dist} with parameters ($\gamma$,$\tau$) taken as $(1.2,1\rm{s})$ and $(3,1\rm{s})$ for the persistence time distributions. \textbf{(C)}-\textbf{(D)} Comparison of concentration profile from the 3D model (thin line) with 1D models. The 3D simulation is ran for elliptical (A) and rectangular (B) channels with $|w_E(0,0)|= 1.5$, $|w_R(0,0)| = 1.0824$, $D_{\rm rot} = 0.035$ and $R = 20,000$ rods. The mean directional persistence times are then computed from the 3D simulations and used as parameters $\tau_{\rm up}, \tau_{\rm down}$ for exponential upstream and downstream distributions in 1D models. The red dashed line is plotted as the expected plateau height of $\mu/V_{\rm eff}$. In all sub-figures, the concentration is computed as number of active rods per unit length $\delta z$.}
    \label{fig:plateauProfiles}
\end{figure}

In both scenarios, we observe the simulated 1D spatiotemporal concentration profiles to display the same general qualitative behavior: rod concentration plateauing around a constant height, while the front of the active rod population advances upstream (Figure~\ref{fig:plateauProfiles}(A)-(B)). Because the power-law distribution of upstream persistence times does not possess a finite second moment, some of the sampled times may vary largely from the mean of $\tau_{\rm up} = 5$s, resulting in greater dispersion of the active rod population. This can be seen in Figure~\ref{fig:plateauProfiles}(B), where the front of the rod population decays at a slower rate than in the exponential case (Figure~\ref{fig:plateauProfiles}(A)), which in contrast has a finite second moment. At short times, this higher dispersion in the power-law case may result in the appearance of the concentration profile sloping upstream. Nevertheless, for both distributions, as long as rods on average are displaced further upstream than they are downstream ($v_{\rm up}\tau_{\rm up} > v_{\rm down}\tau_{\rm down}$), then rods will swim with a net-progress upstream and we will observe emergence of plateau in the concentration profile. If instead $v_{\rm up}\tau_{\rm up} < v_{\rm down}\tau_{\rm down}$, then the concentration profile quickly decays in the positive $z$ direction, and contamination does not occur.

The concentration profile for $t\to\infty$ can be explicitly derived for the 1D biased random walk model with  persistent time distributions of the decaying exponential type. 
Specifically, we derived explicit formulas for the concentration profiles. This result is formulated below. 

\begin{theorem} \label{prop:1}
Let $c(z,t)$ be the concentration profile of the 1D biased random walk model, described in Section~\ref{sec:modelB}. Assume that both upstream and downstream times are given by \eqref{exponential-distribution}, that is, they are random variables exponentially distributed with means $\lambda_u^{-1}$ and $\lambda_d^{-1}$, respectively. Assume also that $c\equiv 0$ at $t=0$, and rods are injected at $z=0$ such that downstream and upstream initial orientations are equiprobable and the injection rate is $\mu$. Finally,  denote upstream and downstream speeds by $V_u$ and $V_d$, as well as define effective coefficients (cf. \eqref{eq:effective_diffusion} and \eqref{eq:effective_velocity} for the case $V=V_u=V_d$):
\[
D_{\text{eff}}=\dfrac{V_uV_d}{\lambda_u+\lambda_d}\text{ and }
V_{\text{eff}}=\dfrac{\lambda_dV_u-\lambda_uV_d}{\lambda_u+\lambda_d}. 
\]
 Then: 
\begin{itemize}
\item [{(i)}] the concentration profile $c(z,t)$ solves the following PDE: 
\begin{equation}
\begin{array}{l}
(\lambda_u+\lambda_d)^{-1}\partial_t^2 c +\partial_t c +V_{\text{eff}}\partial_zc  +(V_u-V_d)(\lambda_u+\lambda_d)^{-1}\partial_z\partial_t c \\
\hspace{30 pt}
= D_{\text{eff}}\partial_z^2 c+ \dfrac{\mu}{2}(V_u-V_d)(\lambda_u+\lambda_d)^{-1}\delta'(z) +\mu\delta(z).
\end{array}\label{prop:eqn-for-c}
\end{equation}
\item [{(ii)}] the large-time concentration profile $c_\infty(z)=\lim\limits_{t\to\infty}c(z,t)$ can be found explicitly via
\begin{equation}
c_\infty(z) = 
\left\{
\begin{array}{lr}
\dfrac{\mu}{V_{\text{eff}}},& z>0,\\
\mu\left(\dfrac{\gamma }{D_{\text{eff}}}+\dfrac{1}{V_{\text{eff}}}\right)\exp\left\{-\dfrac{V_{\text{eff}}}{D_{\text{eff}}}|z|\right\} , & z<0,
\end{array}
\right.\text{ if }V_{\text{eff}}>0,\label{prop:positive_V_eff}
\end{equation}
\end{itemize}
and 
\begin{equation}
c_\infty(z) = 
\left\{
\begin{array}{lr}
\mu\left(\dfrac{1}{|V_{\text{eff}}|}-\dfrac{\gamma }{D_{\text{eff}}}\right)\exp\left\{-\dfrac{|V_{\text{eff}}|}{D_{\text{eff}}}z\right\} , & z>0,
\\
\dfrac{\mu}{|V_{\text{eff}}|},& z<0,
\end{array}
\right.\text{ if }V_{\text{eff}}<0.\label{prop:negative_V_eff}
\end{equation}
Here, $\gamma=\frac{1}{2}(\lambda_u+\lambda_d)^{-1}(V_u-V_d)$.
\end{theorem}
The proof of this theorem can be found in Appendix~\ref{appendix:exponential}. Formulas \eqref{prop:positive_V_eff} and \eqref{prop:negative_V_eff} rigorously justify that for $z>0$ we observe a plateau (a positive constant describing contamination) for $V_{\text{eff}}>0$ and an exponential decaying slope for $V_\text{eff}<0$.  

\begin{remark}
The result of Theorem~\ref{prop:1} can also be derived explicitly if we accept that the concentration profile $c$ solves the advection-diffusion PDE \eqref{advection-diffusion-equation-for-c}. Using \eqref{formula-for-c} we get 
\begin{equation*}
c(z,t) = \dfrac{\mu}{\sqrt{4\pi D_{\text{eff}}}}\exp\left\{\frac{zV_{\text{eff}}}{2D_{\text{eff}}}\right\}\int\limits_0^t\dfrac{1}{\sqrt{\xi}}\exp\left\{-\frac{z^2}{4D_{\text{eff}}\xi}-\frac{V_{\text{eff}}^2\xi}{4D_{\text{eff}}}\right\}\,\mathrm{d}\xi.
\end{equation*}
To pass to the limit $t\to\infty$, we will use the following formula 
\begin{eqnarray*}
\int\limits_0^{\infty}\dfrac{1}{\sqrt{\xi}}\exp\left\{-\dfrac{a^2}{\xi}-b^2\xi\right\}\,\mathrm{d}\xi = \dfrac{\sqrt{\pi}}{b}e^{-2ab} 
\end{eqnarray*}
valid for $a,b>0$. Consider the case $V_{\text{eff}}>0$ (the other case follows similarly). We get 
\begin{equation*}
\lim\limits_{t\to\infty}c(z,t) = \dfrac{\mu}{\sqrt{4\pi D_{\text{eff}}}}\exp\left\{\frac{zV_{\text{eff}}}{2D_{\text{eff}}}\right\}\cdot \dfrac{\sqrt{4\pi D_\text{eff}}}{V_{\text{eff}}}\exp\left\{-\dfrac{|z|V_{\text{eff}}}{2D_{\text{eff}}}\right\}. 
\end{equation*}
Therefore, 
\begin{equation}\label{eq:profileShape}
\lim\limits_{t\to\infty}c(z,t)=
\left\{
\begin{array}{ll}
\dfrac{\mu}{V_{\text{eff}}}\exp\left\{-\dfrac{V_{\text{eff}}}{D_{\text{eff}}} |z|\right\},&z<0,\\ \\
\dfrac{\mu}{V_{\text{eff}}}, & z>0.
\end{array}
\right.
\end{equation}
This formula also captures the behavior illustrated in Figure~\ref{fig:plateauProfiles}: we have a plateau (a constant-in-$z$ solution) for $z>0$ and a decaying exponent for $z<0$ when $V_{\text{eff}}>0$. If $V_{\text{eff}}<0$, the roles of regions $z<0$ and $z>0$ are reversed. In other words, the profile shape for $z>0$, whether it is monotonically decaying or a plateau, depends on the sign of $V_{\text{eff}}$. We note also that \eqref{eq:profileShape} can be considered as a steady state of \eqref{advection-diffusion-equation-for-c}, that is, $V_{\text{eff}}\partial_z c = D_{\text{eff}}\partial_z^2 c+\mu \delta(z)$. 
\end{remark}

Note that the expression for $V_{\rm eff}$, given in Eq.~\eqref{eq:effective_velocity}, is in terms of the mean directional persistence times $\tau_{\rm up}$ and $\tau_{\rm down}$. Moreover, comparison of the simulated concentration profiles with the derived profile shape in Eqs.~\eqref{prop:positive_V_eff},\eqref{prop:negative_V_eff}, and \eqref{eq:profileShape} requires the existence of first moments of the probability distributions for rods' persistence times. For the power-law distribution case, these expected values exist whenever $\gamma > 1$. If instead $\gamma \leq 1$ so that the first moment diverges, then we cannot write $V_{\rm eff}$ as in Eq.~\eqref{eq:effective_velocity} and we can no longer directly compare simulated concentration profiles with the expression for profile shape in Eq.~\eqref{eq:profileShape}. Yet, even in this most extreme case of a divergent first moment, we still observe the same macroscopic behavior: plateau in the concentration profile with a front that advances upstream (shown in Figure~\ref{fig:plateauProfiles}).

We also simulate contamination dynamics in the 3D microchannel under the continuous injection of active rods at $z = 0$. Under this setting, experiments in \cite{figueroa2020coli} reported that concentration profiles of the contamination process was composed of heavy, sloping tails that advance upstream due to the presence of fewer extremely persistent upstream swimmers. Here, however, the simulated concentration profiles of rods along the $z$-axis from the 3D model of active rods in a cylindrical microchannel clearly shows the emergence of a plateau -- i.e., a constant concentration of rods around the expected height of $\mu/V_{\rm eff}$ while a steep front composed of fewer active rods progresses upstream (Figure~\ref{fig:plateauProfiles}(C)-(D)). This profile shape is observed regardless of the channel's cross-sectional geometry. 

\begin{figure}[t!]
    \centering
    \includegraphics[width=0.95\linewidth]{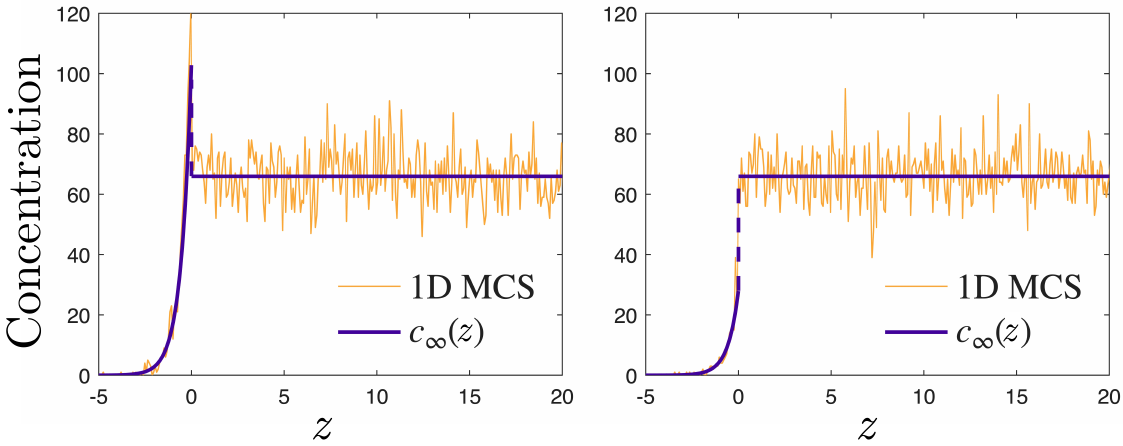}
    \caption{\textbf{Comparison of Monte Carlos simulations for the 1D biased random walk model with analytical expressions for the long-time concentration profile, $\bm{c_\infty (z)}
    $.} Each panel shows 1D Monte Carlo simulations according to the biased random  walk model (1D MCS) compared to the formula Eq.~\eqref{positive_V_eff} for $c_\infty$. On the left, all rods are initiated with upstream orientation, corresponding to $\theta_u = 1, \theta_d = 0$. On the right, rods are initiated facing downstream, $\theta_u = 0, \theta_d = 1$. All other parameters are held constant between the two cases. Concentration is again computed as the total number of active rods counted per unit length.}
    \label{fig:C_infinity}
\end{figure}

Finally, to test the utility of the 1D models for approximating the contamination process observed in 3D, we compare simulated concentration profiles from all three models. As discussed in the previous section, Figure~\ref{fig:histograms} shows that rods' upstream persistence times follow an exponential probability distribution. Thus, the mean directional persistence times $\tau_{\rm{up}}$ and $\tau_{\rm{down}}$ from the 3D simulations were used in 1D models (biased random walk and advection-diffusion) as parameters of the exponential probability distribution (Eq.~\eqref{eq:exp-dist}) for rods' swimming times, $\psi_{\rm{up}}$ and $\psi_{\rm{down}}$. The resulting simulated concentration profiles show that the 1D models closely approximate the overall transport of the active rod population swimming inside microchannels of both rectangular and elliptical cross-section (compared in Figure~\ref{fig:plateauProfiles}(C)-(D)). 

\section{Conclusion}




In this work, we employ a 3D model for active rods swimming in a microchannel to study the physical mechanisms facilitating their upstream transport. Monte Carlo simulations of the 3D model demonstrate upstream contamination occurring for a wide variety of background flow rates, as well as the existence of a flow rate at which rods' net-upstream progress is maximal. By considering microchannels of both elliptic and rectangular cross-section, we also studied the role of cross-sectional geometry on active rods' upstream swimming. Namely, rectangular channels were shown to be more conducive to rods' upstream transport than elliptical channels for a large range of flow rates. Varying channels' aspect ratios demonstrated that rods generally progressed upstream at a higher rate in both rectangular and elliptical channels with a highly narrow cross-section as opposed to channels with cross-sections of equal aspect ratio. These findings highlight the role of confinement on active swimming and also suggest potential geometric controls for mitigating upstream contamination of bacteria such as \textit{E. coli} in medical catheter devices. 

By considering a simplified, 1D biased random walk model of active rods' trajectories, we found that the long-term qualitative behavior of the spatiotemporal concentration profiles depend solely on the expected direction of rods' movement (that is, the sign of the effective velocity $V_{\mathrm{eff}}$). If rods on average persist upstream further than they do downstream, then the macroscopic contamination process was characterized by the same general spatiotemporal profile: plateau near a fixed concentration and an invading front composed of fewer rods advancing upstream. We found this profile shape to emerge for both exponential and heavy-tailed power-law persistence times distributions, and it is also witnessed in simulated concentration profiles from the 3D active rod model. This observation suggests that the population-level behavior of active rods witnessed during contamination of microfluidic channels is largely insensitive to quantitative differences in the underlying persistence-time statistics. These findings complement those of \cite{figueroa2020coli}, which reported that bacterial run-time distributions -- arising from the rotational statistics of the flagellar motor -- play an important role in shaping the earlier stages of macroscopic contamination dynamics. Furthermore, in that work, the presence of a small number of extremely persistent bacteria pioneering the contamination process was linked to a broad distribution governing bacterial run-times. In the modeling framework considered here, our 3D simulations of active rods suggests that persistence times before full directional reversal are distributed exponentially, which is consistent with the classical Poissonian framework for run-and-tumble dynamics. 


To recover non-exponential persistence time distributions (e.g., power-law tails as reported in \cite{korobkova2004molecular}), the 3D model of active rods in a cylindrical microchannel can be extended in the future by incorporating L\'evy walks, which have been reported to better describe bacterial motility \cite{ariel2015swarming}. Furthermore, while reorientation is modeled here by rotational diffusion, a more accurate method for modeling the bacterial tumbling mechanism may be needed to study the statistical properties of specific bacteria that contaminate microfluidic channels. Rod-rod and rod-wall hydrodynamic interactions may also influence transport statistics of rods' trajectories, but are not considered in the current 3D model framework. 

Finally, comparisons between 3D and 1D models demonstrate the utility of simplified, one-dimensional mathematical models to provide a computationally efficient yet accurate description of the more complex transport dynamics observed in 3D.  However, while the one-dimensional PDEs \eqref{advection-diffusion-equation-for-c} and \eqref{prop:eqn-for-c}  can still closely capture the macroscopic transport behavior of active rods even when their persistence time distributions are non-exponential, this limiting equation is only formally valid for exponentially distributed times. Future theoretical studies are needed to elucidate the limiting PDE when directional reversals do not follow a Poisson process but instead are power-law distributed, or possibly a combination of both exponential and power-law distributions depending on the direction that the swimmer is facing.  


\appendix
\section*{Appendices}
\addcontentsline{toc}{section}{Appendices}

\section{Probability distribution function for biased random-walk model}
\label{appendix:velocity-jump}
In this appendix section, our focus is on deriving partial differential equations (PDEs) for the active probability distribution function and the concentration profile in the 1D biased random walk model and the proof of Theorem~\ref{prop:1}. Recall that in this model, the active rod moves down the line with velocity $v_d$ ($v_d<0$), and up the line with velocity $v_u$ ($v_u>0$), and may instantaneously change (jump) between the two velocities (i.e., switch directions). This model resembles the 2-state, continuous time random walk (CTRW), also known as a velocity jump process, where the active rod may be in either the upstream velocity state or the downstream velocity state. Our derivations of PDEs for active rod probability distribution function rely on classical works for CTRW and velocity-jump process \cite{weiss1994aspects,othmer1988models,othmer2000diffusion,othmer2002diffusion}. Additionally, we provide the general PDE in the case of exponential persistent-time distributions \eqref{exponential-distribution}, without imposing any relations between upstream and downstream speeds as well as decay rates of persistence-time distributions. Also, we derive the analytical expressions for limiting concentration profiles. Finally, we consider a mixed random-deterministic case when the downstream time is fixed, as it was considered for validation of experimental results in \cite{figueroa2020coli}.      


 In what follows, the upstream state corresponds to motion with velocity $v_u > 0$ and is denoted by the subscript $u$, while the downstream state corresponds to motion with velocity $v_d < 0$ and is denoted by the subscript $d$. To express general relations without specifying whether a quantity refers to the upstream or downstream state, we use the indices $j$ and $\ell$ to denote the two states, such that if $j = u$, then $\ell = d$, and conversely. For example, for arbitrary quantities $Q_u$ and $Q_d$, the equality
\begin{equation}
Q_j = F(Q_j, Q_\ell)
\text{ 
is equivalent to }
\left\{
\begin{array}{l}
Q_u = F(Q_u, Q_d),\\[4pt]
Q_d = F(Q_d, Q_u).
\end{array}
\right.
\label{j_ell_rule}
\end{equation}

 The active rod initializes its movement at time $T_0 = 0$ and the position $z=0$. Let the swimmer initially move upstream with probability $\theta_u$ and downstream with probability $\theta_d = 1 - \theta_u$. Let $\{T_k\}$ be the set of velocity-jump (turnaround) times that occur at random, where the $k$-th velocity jump occurs at $t = T_k$. The persistence times between jumps, $\mathcal{T}_k = T_k-T_{k-1}$, are independently and identically distributed according to probability density function $\psi_j(t)$ for the corresponding state, that is, 
 \begin{equation}\label{wait_time_density}
     \psi_j(t)\,\text{d} t = \mathbb P\{t\leq \mathcal{T} \leq t +\text{d} t\} \text{ for small }\text{d} t.
 \end{equation}
The function $\Psi_j(t)$ denotes the complement of the cumulative distribution function of $\psi_j$,
\begin{equation}\label{wait_time_ccdf}
     \Psi_j(t) = \mathbb P\{\mathcal{T}\geq t\} = 1-\int\limits^{t}_0\psi_j(\tau)\text{d}\tau.
\end{equation}
\noindent That is, the persistence time of the active rod jumping out of the $j$-th state will be greater than  $t$ amount of time. As a word of caution, the reader should avoid conflating the velocity jump time $T_k$, the moment at which the $k^{\text{th}}$ velocity jump occurs, measured from the initial time, with the persistence time $\mathcal{T}_k = T_k - T_{k-1}$ between the $(k-1)^{\text{th}}$ and $k^{\text{th}}$ jumps. 
The function $\psi_j$ refers to the probability density of the persistence times $\mathcal{T}_k = T_k - T_{k-1}$, as in Section~\ref{sec:modelB}. The subindex $j$ in $\psi_j$ indicates the active rod’s direction of motion during the interval $[T_{k-1}, T_k]$: $j = u$ for upstream motion and $j = d$ for downstream motion.

Introduce $J^{(k)}_j(z,t)$, the conditional probability of that the active rod starts moving with velocity $v_j$ at position $z$ if the $k^{\text{th}}$ velocity jump occurs at time $t$. Observe that the $k^{\text{th}}$ velocity jump may only occur at $(z,t)$ if the $(k-1)^{\text{th}}$ jump occurs at some time $\tau<t$ at some distance $v_\ell(t-\tau)$ away. Therefore, 
\begin{equation}\label{velocity_probability_kth_jump}
    J^{(k)}_j(z,t) = \int\limits^t_0 \psi_\ell(t-\tau)J^{(k-1)}_\ell(z - v_\ell(t-\tau),\tau)\,\text{d}\tau.
\end{equation}
 Summing \eqref{velocity_probability_kth_jump} over all velocity jumps $k$ gives a Volterra integral equation describing the probability of the active rod moving with velocity $v_j$ at position $z$ if any velocity jump occurs at time~$t$,
\begin{equation}\label{velocity_probability_sum_volterra}
\begin{split}
    J_j(z,t) & := \sum\limits^{\infty}_{k=0}J^{(k)}_j(z,t) \\
   & \mathop{=}^{\eqref{velocity_probability_kth_jump}} J^{(0)}_j(z - v_j t ,t) + \int\limits^t_0 \psi_\ell(t-\tau)J_\ell(z - v_\ell(t-\tau),\tau)\,\text{d}\tau.
\end{split}
\end{equation}
Expressing initial conditions as $J_j^{(0)}(z,0) = \theta_j\delta(z)$, where $\theta_j$ is the probability of initializing with velocity $v_j$,
 we rewrite equation \eqref{velocity_probability_sum_volterra} as 

\begin{equation}\label{velocity_probability_volterra}
    J_j(z,t) = \theta_j\delta(z-v_jt)\delta(t)  + \int\limits^t_0 \psi_\ell(t-\tau)J_\ell(z - v_\ell(t-\tau),\tau)\,\text{d}\tau.
\end{equation}
We introduce the density function $p_j(z,t)$, which describes the probability of finding the active rod moving with velocity $v_j$ at position $z$ and time $t$. Equivalently, $p_j(z,t)$ represents the probability that the active rod last jumped to velocity $v_j$ at some time $\tau \le t$, has since traveled a distance $v_j(t - \tau)$, and has a persistence time $\mathcal{T} \ge t - \tau$. Hence, we have the following expression for $p_j(z,t)$:
\begin{equation}\label{general_density_function}
    p_j(z,t) = \int\limits_0^t \Psi_j(t - \tau)\,
    J_j\big(z - v_j(t - \tau), \tau\big)\,\mathrm{d}\tau.
\end{equation}

Next, using the subindex rule \eqref{j_ell_rule}, we write that the general two-state velocity jump process is described by the following equations
\begin{equation}\label{general_two_state_density_system}
    \begin{cases}
        p_u(z,t) = \int\limits^t_0 \Psi_u(t-\tau)J_u(z - v_u(t-\tau),\tau)\,\text{d}\tau,\\
        p_d(z,t) = \int\limits^t_0 \Psi_d(t-\tau)J_d(z - v_d(t-\tau),\tau)\,\text{d}\tau,
    \end{cases}
\end{equation}
with $J_u$ and $J_d$ to be determined from the following system 
\begin{equation}\label{general_two_state_jump_system}
    \begin{cases}
        J_u(z,t) = \theta_u\delta(z-v_ut)\delta(t)  + \int\limits^t_0 \psi_d(t-\tau)J_d(z - v_u(t-\tau),\tau)\,\text{d}\tau,\\
        J_d(z,t) = \theta_d\delta(z-v_dt)\delta(t)  + \int\limits^t_0 \psi_u(t-\tau)J_u(z - v_d(t-\tau),\tau)\,\text{d}\tau.
    \end{cases}
\end{equation}
The total probability density function for finding a swimmer at position $z$ and time $t$ is given by $p(z,t) = p_u(z,t) + p_d(z,t)$.



\subsection{Solution via Fourier-Laplace Transform}

To solve \eqref{velocity_probability_volterra} (or, equivalently, the system \eqref{general_two_state_jump_system}), we apply the Fourier transform in the spatial variable $z$ and the Laplace transform in the temporal variable $t$.
Below, hats denote Fourier transforms and bars denote Laplace transforms, so that, for example, $\hat{J}_j$ and $\hat{\bar{J}}_j$ represent the Fourier and Fourier–Laplace transforms of $J_j$, respectively. Also, $\omega$ and $s$ denote the corresponding Fourier and Laplace variables.

First, taking the Fourier transform of \eqref{velocity_probability_volterra} and using standard properties of the Fourier transform, we get:  

\begin{equation}\label{general_fourier_volterra}
    \hat{J}_j(\omega,t) = \theta_j e^{-i\omega v_jt}\delta(t)  + \int\limits^t_0 \psi_\ell(t-\tau)e^{-i\omega v_\ell(t-\tau)}\hat{J}_\ell(\omega,\tau)\,\text{d}\tau.
\end{equation}
Next, taking the Laplace transform and using its standard properties, we have 
\begin{equation}\label{general_FL_volterra}
    \hat{\bar{J}}_j(\omega,s) = \theta_j  + \bar{\psi}_\ell(s+i\omega v_\ell)\hat{\bar{J}}_\ell(\omega,s).
\end{equation}
This equality can be rewritten as a system 
\begin{equation}\label{general_FL_two_state_jump_system}
    \begin{cases}
        \hat{\bar{J}}_u(\omega,s) = \theta_u  + \bar{\psi}_d(s+i\omega v_d)\hat{\bar{J}}_d(\omega,s),\\
        \hat{\bar{J}}_d(\omega,s) = \theta_d  + \bar{\psi}_u(s+i\omega v_u)\hat{\bar{J}}_u(\omega,s).
    \end{cases}
\end{equation}
Finally, solving this system of two algebraic equations with respect to unknown $\hat{\bar{J}}_u$ and $\hat{\bar{J}}_d$ (or, equivalently, substituting \eqref{general_FL_volterra} with $j$ and $\ell$ swapped into the original form of \eqref{general_FL_volterra} and solving the resulting equality for $\hat{\bar{J}}_j$), we get
\begin{equation}\label{expression_for_J_j}
\hat{\bar{J}}_j(\omega,s) = \dfrac{\theta_j + \theta_\ell \bar{\psi}_\ell(s+i\omega v_\ell)}{1-\bar{\psi}_\ell(s+i\omega v_\ell)\bar{\psi}_j(s+i\omega v_j)}.
\end{equation}
In the same way, the Fourier-Laplace transform of equation \eqref{general_density_function} is
\begin{equation}\label{general_FL_density_function}
    \hat{\bar{p}}_j(\omega,s) = \bar{\Psi}_j(s+i\omega v_j)\hat{\bar{J}}_j(\omega,s).
\end{equation}
From equation \eqref{wait_time_ccdf} we also get the expression for $\bar{\Psi}_j(s)$ in terms of $\bar{\psi}_j(s)$:
\begin{equation}\label{wait_time_ccdf_FL}
    \bar{\Psi}_j(s) = \dfrac{1-\bar{\psi}_j(s)}{s}.
\end{equation}
Collecting  \eqref{expression_for_J_j},  \eqref{general_FL_density_function}, and \eqref{wait_time_ccdf_FL}, we find the expression for the Fourier-Laplace transform of the probability density for finding a swimmer at position $z$ and time $t$: 
\begin{multline}\label{general_FL_total_density}
    \hat{\bar{p}}(\omega,s) = \frac{1}{1-\bar{\psi}_d(s+i\omega v_d)\bar{\psi}_u(s+i\omega v_u)}\times\\
    \times\bigg\{\frac{(1-\bar{\psi}_u(s+i\omega v_u))(\theta_u  + \theta_d\bar{\psi}_d(s+i\omega v_d))}{s+i\omega v_u}+\\
    +\frac{(1-\bar{\psi}_d(s+i\omega v_d))(\theta_d  + \theta_u\bar{\psi}_u(s+i\omega v_u))}{s+i\omega v_d} \bigg\}.
\end{multline}
Note that the expression above is the analytical solution for the Fourier-Laplace transform of the probability distribution function $p(z,t)$, given the persistence-time distributions $\psi_u$ and $\psi_d$. By taking the inverse Laplace transform, with the account of initial conditions   
$$p_j(z,0)=\theta_j\delta(z),$$ one can find the Fourier transform of the total distribution $p(z,t)$. We will show how the cumbersome expression \eqref{general_FL_density_function} can be reduced to partial differential equations in specific cases of the persistence-time distributions $\psi_u$ and $\psi_d$.

We complete the presentation of the solution in the general case of the persistence-time distributions $\psi_u$ and $\psi_d$ with the expression for $p_{u}(z,t)$ and $p_{d}(z,t)$ describing their dynamics as the combination of the advection with their respective velocities, $v_{u}$ and $v_{d}$, together with coupling between upstream and downstream swimming fluxes. Specifically, substitute \eqref{wait_time_ccdf_FL} into \eqref{general_FL_density_function} and multiply both sides of the resulting equality by $s+i\omega v_j$: 
\begin{equation*}
s\hat{\bar{p}}_j+i\omega v_j\hat{\bar{p}}_j = \hat{\bar{J}}_j - \bar{\psi}_j(s+i\omega v_j) \hat{\bar{J}}_j. 
\end{equation*}
Next, use \eqref{general_FL_volterra} to rewrite the first term in the right-hand side of the equality above and then subtract $\theta_j$ from both sides: 
\begin{equation}\label{general_system_FL}
s\hat{\bar{p}}_j-\theta_j+i\omega v_j\hat{\bar{p}}_j = \bar{\psi}_\ell(s+i\omega v_\ell)\hat{\bar{J}}_\ell - \bar{\psi}_j(s+i\omega v_j) \hat{\bar{J}}_j. 
\end{equation}
Here, the argument of $\hat{\bar{p}}_j$, $\hat{\bar{J}}_\ell$, $\hat{\bar{J}}_j$ is $(\omega,s)$. Now, use that the inverse Fourier-Laplace transform of $s\hat{\bar{p}}_j-\theta_j$ and $i\omega v_j\hat{\bar{p}}_j$ turn into $\partial_t p_j$ and $v_j\partial_zp_j$, respectively. Next, shift and product properties of the Fourier and Laplace transforms applied to the right-hand side of \eqref{general_system_FL} imply:
\begin{equation}\label{general_advection_equation_with_coupling}
    \partial_tp_j+v_j\partial_zp_j =\int\limits_0^t \left[\psi_{\ell}(\tau)J_{\ell}(z-v_{\ell}\tau,t-\tau) - \psi_j(\tau)J_j(z-v_j\tau,t-\tau)\right]\,\text{d}\tau.
\end{equation}
This equation writes as a system for $p_u$ and $p_d$ as follows:
\begin{equation}\label{general_advection_system_with_coupling}
    \hspace{0 pt}\begin{cases}
     \partial_tp_u+v_u\partial_zp_u =\int\limits_0^t \psi_{d}(\tau)J_{d}(z-v_{d}\tau,t-\tau) - \psi_u(\tau)J_u(z-v_u\tau,t-\tau)\,\text{d}\tau, \\
     \partial_tp_d+v_d\partial_zp_d =\int\limits_0^t \psi_{u}(\tau)J_{u}(z-v_{u}\tau,t-\tau) - \psi_d(\tau)J_d(z-v_d\tau,t-\tau)\,\text{d}\tau.
    \end{cases}
\end{equation}
Note that the system \eqref{general_advection_system_with_coupling} implies that 
\begin{equation}\label{formula_for_time_derivative}
\partial_t p = - \partial_z J, \quad \text{where }J:=v_up_u+v_dp_d.
\end{equation}

\subsection{Reduction to PDE in the case of exponential persistence-time distributions}
\label{appendix:exponential}
Consider the exponential persistence-time distribution $\psi_j(t) = \lambda_j e^{-\lambda_j t}$, that is, 
\begin{equation*}
\psi_u(t) = \lambda_u e^{-\lambda_u t}\text{ and }\psi_d(t) = \lambda_d e^{-\lambda_d t}
\end{equation*}
for some $\lambda_u,\lambda_d>0$. In this case, the Laplace transforms of $\psi_j$ and $\Psi_j$ are: 
\begin{equation}\label{exponential_laplace}
\bar{\psi}_j(s) = \dfrac{\lambda_j}{s+\lambda_j} \text{ and }\bar{\Psi}_j(s)=\dfrac{1}{s+\lambda_j}. 
\end{equation}
As a consequence, we have the following relation between $\bar{\psi}_j(s)$ and $\bar{\Psi}_j(s)$: 
\begin{equation}
\label{exponential_expression}
\bar{\psi}_j(s)=\lambda_j\bar{\Psi}_j(s).
\end{equation}
We note here that due to \eqref{wait_time_ccdf_FL}, the relation 
holds only for the exponential persistence-time distributions.  

Substituting \eqref{exponential_expression} into \eqref{general_FL_density_function} we get 
\begin{equation*}
\lambda_j \hat{\bar{p}}_j (\omega,s)= \bar{\psi}_j(s+i\omega v_j)\hat{\bar{J}}_j(\omega,s). 
\end{equation*}
Using this identity to rewrite both terms on the right-hand side of \eqref{general_system_FL}, we get 
\begin{equation*}
s\hat{\bar{p}}_j-\theta_j+i\omega v_j\hat{\bar{p}}_j = \lambda_\ell \hat{\bar{p}}_\ell - \lambda_j \hat{\bar{p}}_j.    
\end{equation*}
Similarly to how we obtained \eqref{general_advection_equation_with_coupling} from \eqref{general_system_FL}, we use the inverse Fourier and Laplace transforms to imply  
\begin{equation*}
\partial_t p_j + v_j \partial_z p_j = \lambda_\ell p_\ell - \lambda_j p_j,
\end{equation*}
which reads as a system for unknowns $p_u$ and $p_d$ as follows
\begin{equation}\label{master_equation_two_direction_system}
    \begin{cases}
    \partial_tp_u +v_u\partial_zp_u = \lambda_d p_d -\lambda_u p_u,\\
    \partial_tp_d +v_d\partial_zp_d = \lambda_u p_u -\lambda_d p_d.
    \end{cases}
\end{equation}
Thus, this one-dimensional velocity jump process in the case of exponential persistence-time distributions reduces to the telegrapher's system \cite{othmer1988models,weiss1994aspects}. Below, in this section we consider several particular cases of \eqref{master_equation_two_direction_system} with the purpose to derive a closed form partial differential equation for the total probability distribution function $p=p_u+p_d$. To this end, we rewrite \eqref{general_FL_total_density} for the case of \eqref{exponential_laplace} as follows: 
\begin{equation}\label{general_formula_for_exponential_case}
\hat{\bar{p}}(\omega,s) = \dfrac{\theta_u m_d + \theta_d m_u+(\lambda_u+\lambda_d)(\theta_u+\theta_d)}{m_dm_u+\lambda_u m_d+\lambda_d m_u},
\end{equation}
where $m_u=s+i\omega v_u$ and $m_d = s+ i\omega v_d$. One can slightly simplify \eqref{general_formula_for_exponential_case} by using  $\theta_u+\theta_d=1$. We keep $\theta_u+\theta_d$ in \eqref{general_formula_for_exponential_case} since we believe it will make rearrangements of terms below clearer. We will also need initial conditions for $\hat{p}(\omega,t)$: 
\begin{equation}\label{initial_condition_for_hat_p}
\begin{array}{l}
\,\,\,\,\,\hat{p}|_{t=0}=\theta_u+\theta_d,\\
\partial_t\hat{p}|_{t=0} = -i\omega (\theta_uv_u+\theta_d v_d).
\end{array}
\end{equation}
Here, we used that $p|_{t=0}=p_u|_{t=0}+p_d|_{t=0}=(\theta_u+\theta_d)\delta(z)$ and 
$$
\partial_tp|_{t=0}=\left.\left(\partial_t p_u+\partial_t p_d\right)\right|_{t=0} \mathop{=}^{\eqref{formula_for_time_derivative}}\left.\left(-v_u \partial_z p_u-v_d \partial_z p_d\right)\right|_{t=0}=-(v_u\theta_u+v_d\theta_d)\delta'(z), 
$$
as well as that Fourier transforms of $\delta(z)$ and $\delta'(z)$ are 1 and $i\omega$, respectively. 

Rewrite \eqref{general_formula_for_exponential_case} so it is convenient to take the inverse Fourier-Laplace transform: 
\begin{eqnarray*}
&& \left[s^2\hat{\bar{p}}-(\theta_u+\theta_d)s + i\omega (v_u\theta_u+v_d\theta_d)\right] \\
&& \hspace{30 pt}
+(\lambda_u+\lambda_d)\left[s\hat{\bar{p}}-(\theta_u+\theta_d)\right]  \\
&& \hspace{30 pt}
+i\omega (\lambda_u v_d+\lambda_d v_u) \hat{\bar{p}}\\
&& \hspace{30 pt}
+i\omega(v_d + v_u)\left[s\hat{\bar{p}}-(\theta_u+\theta_d)\right]  \\
&& \hspace{30 pt} -\omega^2 v_u v_d \hat{\bar{p}} = 0.
\end{eqnarray*}
Using the initial conditions \eqref{initial_condition_for_hat_p} we get a closed form equation for $p(z,t)$:
\begin{equation} \label{exponential_closed_form_for_p}
\begin{array}{l}
\partial_t^2 p +(\lambda_u+\lambda_d)\partial_t p +(\lambda_u v_d + \lambda_d v_u)\partial_zp  \\
\hspace{80 pt}
+(v_u+v_d)\partial_z\partial_t p = - v_uv_d\partial_z^2 p.
\end{array}
\end{equation}

Next, we consider equations for the total probability distribution $p=p_u+p_d$ in particular cases of exponential persistence-time distributions: 

\begin{itemize}
\item[{$-$}] {\it Symmetric case, when the active rod moves with the same speed $V=v_u=-v_d$ upstream and downstream and persistence time rates are the same $\lambda = \lambda_u=\lambda_d$, so $\psi(t):=\psi_u(t)=\psi_d(t) =\lambda e^{-\lambda t}$}. In this case, the telegrapher's system can be reduced to the damped wave equation: 
\begin{equation}
\partial_t^2 p + 2\lambda \partial_t p = V^2 \partial_z^2 p.\label{pde-same-lambda-and-v} 
\end{equation}

\item[{$-$}] {\it Equal persistence time rates $\lambda=\lambda_u = \lambda_d$ but different upstream and downstream speeds, $V_u=v_u$, $V_d = -v_d$, and $V_u\neq V_d$}: 
\begin{equation}\label{pde_neq_speed_poisson}
    \partial^2_tp + 2\lambda \partial_tp +(V_u-V_d)\partial_z[\partial_tp+\lambda p] = V_uV_d\partial^2_zp. 
\end{equation}

\item[{$-$}] {\it Equal speeds $V=v_u=-v_d$ but different rates $\lambda_u\neq \lambda_d$}:
\begin{equation}\label{pde_neq_lambda_poisson}
    \partial_t^2p + (\lambda_u+\lambda_d)\partial_tp+V(\lambda_d-\lambda_u)\partial_zp=V^2\partial_z^2p.
\end{equation}

\end{itemize}

\smallskip 

\noindent{\it Nondimensionalization and formulas for effective drift $V_{\mathrm{eff}}$ and diffusion $D_{\mathrm{eff}}$}. Next, we discuss a relation between this equation and the advection-diffusion equation in Model C. Specifically, in a non-dimensional form for time scale $T$ and spatial scale $L$ with 
\begin{equation*}
\hat{t} = \dfrac{t}{T}, \quad \hat{z}= \dfrac{z}{L}, \quad 
\lambda_j = \dfrac{\hat{\lambda}_j}{T},\quad v_j = \hat{v}_j\dfrac{L}{T},
\end{equation*}
and dividing both sides by $\hat{\lambda}_u+\hat{\lambda}_d$, the equation \eqref{exponential_closed_form_for_p} becomes
\begin{equation} \label{exponential_closed_form_for_p_nondim}
\begin{array}{l}
\dfrac{1}{\hat{\lambda}_u+\hat{\lambda}_d}\partial_{\hat{t}}^2 p +\partial_{\hat{t}} p +\dfrac{\hat{\lambda}_u \hat{v}_d + \hat{\lambda}_d \hat{v}_u}{\hat{\lambda}_u+\hat{\lambda}_d}\partial_{\hat{z}}p  \\
\hspace{80 pt}
+\dfrac{\hat{v}_u+\hat{v}_d}{\hat{\lambda}_u+\hat{\lambda}_d}\partial_{\hat{z}}\partial_{\hat{t}} p = - \dfrac{\hat{v}_u\hat{v}_d}{\hat{\lambda}_u+\hat{\lambda}_d}\partial_{\hat{z}}^2 p.
\end{array}
\end{equation}
If we drop the first and fourth terms on the left-hand side of \eqref{exponential_closed_form_for_p_nondim}, which would be reasonable for small average waiting times $\hat{\lambda}_{u}^{-1},\hat{\lambda}_{d}^{-1}\ll 1$, and in view that velocities $\hat{v}_u$ and $\hat{v}_d$ have opposite signs, $v_uv_d<0$, we will arrive to the following advection-diffusion equation: 
\begin{equation}\label{advection-diffusion-equation-appendix}
\partial_{\hat{t}} p +\hat{V}_{\text{eff}}\,\partial_{\hat{z}}p 
 = \hat{D}_{\text{eff}}\,\partial_{\hat{z}}^2 p,
\end{equation}
where 
\begin{equation}\label{effective_velocity_and_diffusion}
\hat{V}_{\text{eff}} = \dfrac{\hat{\lambda}_u \hat{v}_d + \hat{\lambda}_d \hat{v}_u}{\hat{\lambda}_u+\hat{\lambda}_d} \text{ and }\hat{D}_{\text{eff}}= \dfrac{-\hat{v}_u\hat{v}_d}{\hat{\lambda}_u+\hat{\lambda}_d}.
\end{equation}
We also write \eqref{effective_velocity_and_diffusion} in terms of average persistence times $\hat{\tau}_j = \hat{\lambda}_j^{-1}$ instead of parameters~$\lambda_j$:
\begin{equation}\label{tau:effective_velocity_and_diffusion}
\hat{V}_{\text{eff}} = \dfrac{ \hat{v}_u\hat{\tau}_u +  \hat{v}_d\hat{\tau}_d}{\hat{\tau}_u+\hat{\tau}_d} \text{ and }\hat{D}_{\text{eff}}= -\hat{v}_u\hat{v}_d\dfrac{\hat{\tau}_u\hat{\tau}_d}{\hat{\tau}_u+\hat{\tau}_d}.
\end{equation}

\smallskip 

\noindent{\it Large-time behavior of swimming concentration $c(z,t)$.} In the contamination scenario, where active rods are injected at $ z = 0 $ with constant frequency $ \mu $, numerical simulations show that the concentration profile converges to a plateau on one side of $z = 0 $ and exhibits exponential decay on the other side, depending on the sign of the effective velocity $V_{\mathrm{eff}}$ defined in \eqref{tau:effective_velocity_and_diffusion}. In the case of exponential persistence-time distributions \eqref{exponential_laplace}, we can derive this limiting profile analytically. 

First, we note that from the formula for the concentration profile $c$ \eqref{c_via_p} and the equation for $p$ \eqref{exponential_closed_form_for_p}, we get the equation for the concentration profile $c$: 
\begin{equation} \label{concentration_equation:exponential}
\begin{array}{l}
\partial_t^2 c +(\lambda_u+\lambda_d)\partial_t c +(\lambda_u v_d + \lambda_d v_u)\partial_zc  +(v_u+v_d)\partial_z\partial_t c \\
\hspace{30 pt}
= - v_uv_d\partial_z^2 c+ \mu(\theta_u v_d+\theta_d v_u)\delta'(z) +\mu(\lambda_u+\lambda_d)(\theta_u+\theta_d)\delta(z).
\end{array}
\end{equation}
The equation \eqref{concentration_equation:exponential} is the same as \eqref{exponential_closed_form_for_p} with addition of singular terms in the right-hand side. To derive the non-homogeneous singular terms, we used 
\begin{eqnarray*}
&p=\mu^{-1}\partial_tc, \quad \int\limits_{0}^{t}\partial_t p(z,\tilde{t})\,\text{d}\tilde{t} = \mu^{-1}\partial_tc -(\theta_u+\theta_d)\delta(z),&\\&\,\,\int\limits_0^t \partial_t^2 p(z,\tilde{t})\,\text{d}\tilde{t}=\mu^{-1}\partial_t^2 c +(\theta_u v_u + \theta_d v_d)\delta'(z). &
\end{eqnarray*}

Next, the formula \eqref{c_via_p} also implies that 
\begin{equation}\nonumber
\hat{\bar{c}} = \mu \dfrac{\hat{\bar{p}}}{s}.
\end{equation}
Then, formal application of the finite value theorem leads to 
\begin{eqnarray}
\hat{c}_{\infty}(\omega):=\lim\limits_{t\to\infty}\hat{c}(\omega, t) &=& \lim\limits_{s\to 0 }s\cdot \left(\mu \dfrac{\hat{\bar{p}}(\omega,s)}{s}\right)\nonumber\\ 
&=& \mu \lim\limits_{s\to 0 }\hat{\bar{p}}(\omega,s) \nonumber\\
&=& \mu \dfrac{(\theta_u v_d +\theta_d v_u)i\omega + (\lambda_u+\lambda_d)(\theta_u+\theta_d)}{-v_dv_u \omega^2 + (\lambda_u v_d + \lambda_d v_u)i\omega}. \label{hat_c_infty}
\end{eqnarray}
Note that this formula corresponds to that the limiting concentration profile $c_\infty (z)$ is the steady state of \eqref{concentration_equation:exponential}, thus, satisfying the following differential equation:
\begin{equation}\label{steady_state_c_infty}
(\lambda_u v_d + \lambda_d v_u)\partial_zc_\infty=(- v_uv_d)\partial_z^2 c_\infty+ \mu(\theta_u v_d+\theta_d v_u)\delta'(z) +\mu(\lambda_u+\lambda_d)(\theta_u+\theta_d)\delta(z).
\end{equation}

Similar to formulas in \eqref{effective_velocity_and_diffusion}, define $V_{\text{eff}}:=(\lambda_uv_d+\lambda_dv_u)/(\lambda_u+\lambda_d)$ and $D_{\text{eff}}:=(-v_dv_u)/(\lambda_u+\lambda_d)$. Introduce also $\gamma:=(\theta_uv_d+\theta_d v_u)/(\lambda_u+\lambda_d)$. Then the formula for $\hat{c}_{\infty}$ becomes 
\begin{equation}\label{as_fraction_hat_c_infty}
    \hat{c}_{\infty}(\omega)=\mu\dfrac{\gamma i\omega + (\theta_u+\theta_d)}{D_{\text{eff}}\omega^2+V_{\text{eff}}i\omega},
\end{equation}
or, if we multiply both sides of the equation by the denominator in the right-hand side, 
\begin{equation}\label{fourier_equation_c_infty}
\left[D_{\text{eff}}\omega^2+V_{\text{eff}}i\omega\right]\hat{c}_{\infty}(\omega)=\mu\left(\gamma i\omega + (\theta_u+\theta_d)\right),
\end{equation}
which is the Fourier transform of the equation \eqref{steady_state_c_infty}. Here, it is important to note that if $\hat{c}_{\infty}(\omega)$ solves \eqref{fourier_equation_c_infty}, then 
\begin{equation*}
\hat{c}_{\infty}(\omega) + A \delta(\omega) + B \delta(\omega + i(V_{\text{eff}}/D_{\text{eff}}))
\end{equation*}
is also a solution of \eqref{fourier_equation_c_infty} for any constants $A$ and $B$. It is consistent with the general solution of the equation \eqref{steady_state_c_infty} which is the sum of a particular solution $c_{\infty}(z)$, obtained, for example, by the inverse Fourier transform of \eqref{as_fraction_hat_c_infty} and a linear combination of particular solutions of the homogeneous version of \eqref{steady_state_c_infty}, $A+Be^{(V_{\text{eff}}/D_{\text{eff}})z}$. We find a particular solution by performing partial fraction decomposition for the right-hand side of \eqref{as_fraction_hat_c_infty} as follows: 
\begin{equation}
\mu\dfrac{\gamma i\omega + (\theta_u+\theta_d)}{D_{\text{eff}}\omega^2+V_{\text{eff}}i\omega}=\dfrac{\mu (\theta_u+\theta_d)}{V_{\text{eff}}}\dfrac{1}{i\omega}-\mu\left(\dfrac{\gamma }{D_{\text{eff}}}+\dfrac{(\theta_u+\theta_d)}{V_{\text{eff}}}\right)\dfrac{1}{i\omega - (V_{\text{eff}}/D_{\text{eff}})}.
\end{equation}
By using standard inverse Fourier transforms 
\begin{equation*}
    \mathcal{F}^{-1}\left[\dfrac{1}{i\omega}\right]=\dfrac{1}{2}\text{sgn}(z)\text{ and }\mathcal{F}^{-1}\left[\dfrac{1}{i\omega-a}\right]=\left\{
    \begin{array}{ll}
    -e^{az}H(-z),& a>0,\\
    e^{az}H(z), & a<0,
    \end{array}
    \right.
\end{equation*}
where $H(z)$ is the Heaviside function and $\text{sgn}(z)=2H(z)-1$ is the sign function, and choosing $A$ and $B$ so that $c_{\infty}$ is positive, bounded, and decays to zero either at $z\to \infty$ or $z\to -\infty$ we  get {\it the analytical expression for the large time concentration profile} $c_{\infty}$: 
\begin{equation}
c_\infty(z) = 
\left\{
\begin{array}{lr}
\dfrac{\mu(\theta_u+\theta_d)}{V_{\text{eff}}},& z>0,\\
\mu\left(\dfrac{\gamma }{D_{\text{eff}}}+\dfrac{(\theta_u+\theta_d)}{V_{\text{eff}}}\right)e^{V_{\text{eff}}/D_{\text{eff}}}z , & z<0,
\end{array}
\right.\label{positive_V_eff}
\end{equation}
for the case $V_{\text{eff}}>0$, and 
\begin{equation}
c_\infty(z) = 
\left\{
\begin{array}{lr}
\mu\left(\dfrac{(\theta_u+\theta_d)}{|V_{\text{eff}}|}-\dfrac{\gamma }{D_{\text{eff}}}\right)e^{-|V_{\text{eff}}|/D_{\text{eff}}}z , & z>0,
\\
\dfrac{\mu(\theta_u+\theta_d)}{|V_{\text{eff}}|},& z<0,
\end{array}
\right.\label{negative_V_eff}
\end{equation}
for the case $V_{\text{eff}}<0$. Therefore, for exponentially decaying persistence time distributions, we analytically confirm that, depending on the sign of $V_{\mathrm{eff}}$, a plateau is observed for either $ z > 0 $ or $ z < 0 $, while the profile decays exponentially for $ z < 0 $ and $ z > 0 $, respectively. Derivation of formulas \eqref{positive_V_eff} and \eqref{negative_V_eff} concludes the proof of Theorem~\ref{prop:1}.

When $V_{\rm eff} = 0$, no non-trivial steady state $c_\infty(z)$ exists. The swimmer population instead grows unboundedly near $z=0$ while slowly spreading toward $z \to \pm\infty$.

\smallskip 

\subsection{Reduction to PDE in the case of exponential upstream per\-sis\-tence-time distribution $\psi_u(t)=\lambda_u e^{-\lambda_u t}$ and deterministic downstream persistence time $\psi_d(t)= \delta(t-T_d)$.}
\label{appendix:deterministic}

Here, we will consider the case related to the model introduced in \cite{figueroa2020coli} for the bacterial profiles spreading in microchannels. In this model, the downstream persistence time is assumed to be deterministic, that is, $\psi_d(t)=\delta(t-T_d)$ for some $T_d>0$. Also, we take the exponential distribution for the upstream distribution function $\psi_u(t)=\lambda_u e^{-\lambda_u t}$. Such a choice of persistence-time distributions represents the mixed case of both random and deterministic turnarounds. Laplace transform of persistence-time distributions are 
\begin{equation*}
\bar{\psi}_u(s) = \dfrac{\lambda_u}{s+\lambda_u} \quad\text{and}\quad \bar{\psi}_d(s) = e^{-T_ds}. 
\end{equation*}
Substituting these expressions in \eqref{general_FL_total_density} we get 
\begin{equation}\label{FL_p_for_delta_case}
\hat{\bar{p}}=\dfrac{(\theta_u+\theta_d e^{-T_dm_d})m_d+(1-e^{-T_dm_d})(\theta_d(m_u+\lambda_u)+\theta_u\lambda_u)}{m_d(m_u+\lambda_u-\lambda_ue^{-m_dT_d})},
\end{equation}
where $m_u=s+i\omega v_u$ and $m_d = s+ i\omega v_d$. We note here that if one expands the expression for $\bar{\psi}_d(s)$ for small $s$ and denotes $\lambda_d := T_{d}^{-1}$, then $\bar{\psi}_d=1/e^{T_ds}\approx \lambda_d/(s+\lambda_d)$ and \eqref{FL_p_for_delta_case} transforms into the formula \eqref{general_formula_for_exponential_case} for the exponential case allowing for a further simplification to the advection-diffusion equation \eqref{advection-diffusion-equation-appendix}.

Multiplying both sides of \eqref{FL_p_for_delta_case} by the denominator of the right-hand side and rearranging terms we obtain: 
\begin{eqnarray*}
&&\left[s^2\hat{\bar{p}}-(\theta_u+\theta_d)s+i\omega(v_u\theta_u+v_d\theta_d)\right]\\
&&\hspace{30pt}+\left\{i\omega(v_d+v_u)+\lambda_u\right\}\left[s\hat{\bar{p}}-(\theta_u+\theta_d)\right]\\
&&\hspace{30pt}+\left\{-v_dv_u\omega^2 +\lambda_uv_d(i\omega)\right\}\hat{\bar{p}}\\
&&\hspace{30pt}-\lambda_u e^{-v_dT_d(i\omega)}e^{-sT_d}\left[s\hat{\bar{p}}-(\theta_u+\theta_d)\right]\\
&&\hspace{30pt}-\lambda_uv_d(i\omega)e^{-v_dT_d(i\omega)}e^{-sT_d}\hat{\bar{p}}=e^{-v_dT_d(i\omega)}e^{-sT_d}(i\omega) \theta_d(v_d-v_u). 
\end{eqnarray*}
First, taking the inverse Laplace transform with the account of initial conditions \eqref{initial_condition_for_hat_p} and then the inverse Fourier transform, we get: 
\begin{eqnarray}\label{eqn_for_p_in_deterministic_case}
&& \partial_t^2 p+(v_d+v_u)\partial_z\partial_t p +\lambda_u\partial_tp+v_dv_u\partial_z^2 p +\lambda_u v_d \partial_z p\nonumber\\&& \hspace{40pt} -\lambda_u H(t-T_d) \left(\partial_tp(z-v_dT_d,t-T_d)+v_d \partial_zp(z-v_dT_d,t-T_d)\right)\nonumber\\
&&\hspace{30pt}=\theta_d(v_d-v_u)\delta(t-T_d)\delta'(z-v_dT_d).
\end{eqnarray}

\noindent Here, the singular term in the right-hand side is due to active rods injected at $t=0$ with downstream orientations. These active rods travel from $z=0$ to $z=v_dT_d$ during the time interval $0<t<T_d$. If we assume that all active rods are initiated with upward direction, that is, $\theta_d=0$, the singular term disappears. Another consequence of deterministic downstream persistence-time distribution is memory terms in the left-hand side of \eqref{eqn_for_p_in_deterministic_case}. 

In the same way as we derived \eqref{concentration_equation:exponential}, we write the equation for concentration equation in this case: 
\begin{eqnarray}\label{eqn_for_c_in_deterministic_case}
&& \partial_t^2 c+(v_d+v_u)\partial_z\partial_t c +\lambda_u\partial_tc+v_dv_u\partial_z^2 c +\lambda_u v_d \partial_z c\nonumber\\&& \hspace{40pt} -\lambda_u H(t-T_d) \left(\partial_tc(z-v_dT_d,t-T_d)+v_d \partial_zc(z-v_dT_d,t-T_d)\right)\nonumber\\
&&\hspace{30pt}=\mu\theta_d(v_d-v_u)H(t-T_d)\delta'(z-v_dT_d)+\lambda_u\mu(\theta_u+\theta_d)\delta(z)\nonumber\\
&&\hspace{50pt}+\mu(\theta_uv_d+\theta_d v_u)\delta'(z)-\lambda_u\mu(\theta_u+\theta_d) H(t-T_d)\delta(z-v_dT_d).
\end{eqnarray}
The computation of an explicit analytical formula for the steady state $c_\infty$ in this case is technically cumbersome and is therefore deferred to future work.

\section{Nondimensionalization of the three-dimensional model of the active rod}
\label{appendix:nondim}
Introduce unit of length and unit of time as $L$ and $T$, respectively, so that $\boldsymbol{r}=\hat{\boldsymbol{r}}L$ and $t=\hat{t}T$ and the non-dimensionalized system looks as follows  
\begin{equation}
\left\{
    \begin{array}{l}
   \partial_{\hat{t}}\hat{\boldsymbol{r}} = \dfrac{V_0T}{L}\boldsymbol{p} + \dfrac{wT}{L}\boldsymbol{e}_z,\\
    \text{d}{\boldsymbol{p}} = (\text{I}-\boldsymbol{p}\boldsymbol{p}^{\text{T}})\left[\nabla_{\hat{\boldsymbol{r}}} \left(\dfrac{\boldsymbol{u}T}{L}\right)\boldsymbol{p}\,\text{d}\hat{t} + \sqrt{2D_{\text{rot}}T}\,\text{d}\hat{W}\right].
    \end{array}
    \right.
\end{equation}
So these are non-dimensional parameters (used in numerical simulations): 
\begin{equation}
\hat{V}_{0}=\dfrac{V_0T}{L}\text{ and } \hat{D}_{\text{rot}} = D_{\text{rot}}T.
\end{equation}
Average swimming speed and rotational diffusion for the same bacterial strain can vary depending on macroscopic environmental parameters such as fluid viscosity and nutrient saturation. For example, a representative set of values for {\it Bacillus subtilis} is \cite{haines2009three,sokolov2009reduction,ryan2011viscosity}
\begin{equation}
\ell = 10\, \mu \text{m}, \quad V_0 = 20 \,\mu\text{m}\cdot \text{s}^{-1}, \quad D_{\text{rot}} = 0.1\, \text{s}^{-1}. 
\end{equation}
Choose $L$ and $T$ such that $\hat{\ell}=0.2$ (non-dimensional active rod length) and $\hat{V}_0=1$: 
\begin{equation}
L=\dfrac{\ell}{\hat{\ell}} = 50\, \mu \text{m} \text{ and }T=L\dfrac{\hat{V}_0}{V_0}=2.5\, \text{s}.
\end{equation}
\clearpage
\bibliographystyle{unsrt}
\bibliography{Literature}

@article{figueroa2020coli,
  title={E. coli “super-contaminates” narrow ducts fostered by broad run-time distribution},
  author={Figueroa-Morales, Nuris and Rivera, Aramis and Soto, Rodrigo and Lindner, Anke and Altshuler, Ernesto and Cl{\'e}ment, {\'E}ric},
  journal={Science advances},
  volume={6},
  number={11},
  pages={eaay0155},
  year={2020},
  publisher={American Association for the Advancement of Science}
}

@article{brown2025boundary,
  title={Boundary accumulation of active rods in microchannels with elliptical cross-section},
  author={Brown, Chase and Potomkin, Mykhailo and Ryan, Shawn D},
  journal={Physical Review E},
  volume={112},
  number={4},
  pages={045407},
  year={2025},
  publisher={APS}
}

@article{rivero1989transport,
  title={Transport models for chemotactic cell populations based on individual cell behavior},
  author={Rivero, Mercedes A and Tranquillo, Robert T and Buettner, Helen M and Lauffenburger, Douglas A},
  journal={Chemical engineering science},
  volume={44},
  number={12},
  pages={2881--2897},
  year={1989},
  publisher={Elsevier}
}

@article{ahmed2008experimental,
  title={Experimental verification of the behavioral foundation of bacterial transport parameters using microfluidics},
  author={Ahmed, Tanvir and Stocker, Roman},
  journal={Biophysical journal},
  volume={95},
  number={9},
  pages={4481--4493},
  year={2008},
  publisher={Elsevier}
}

@article{lovely1975statistical,
  title={Statistical measures of bacterial motility and chemotaxis},
  author={Lovely, Peter S and Dahlquist, FW},
  journal={Journal of theoretical biology},
  volume={50},
  number={2},
  pages={477--496},
  year={1975},
  publisher={Elsevier}
}

@article{mallikarjun2023chiral,
  title={Chiral run-and-tumble walker: Transport and optimizing search},
  author={Mallikarjun, Rahul and Pal, Arnab},
  journal={Physica A: Statistical Mechanics and its Applications},
  volume={622},
  pages={128821},
  year={2023},
  publisher={Elsevier}
}

@article{aranson2022bacterial,
  title={Bacterial active matter},
  author={Aranson, Igor S},
  journal={Reports on Progress in Physics},
  volume={85},
  number={7},
  pages={076601},
  year={2022},
  publisher={IOP Publishing}
}

@article{thiel2012anomalous,
  title={Anomalous diffusion in run-and-tumble motion},
  author={Thiel, Felix and Schimansky-Geier, Lutz and Sokolov, Igor M},
  journal={Physical Review E—Statistical, Nonlinear, and Soft Matter Physics},
  volume={86},
  number={2},
  pages={021117},
  year={2012},
  publisher={APS}
}

@article{villa2020run,
  title={Run-and-tumble bacteria slowly approaching the diffusive regime},
  author={Villa-Torrealba, Andrea and Ch{\'a}vez-Raby, Crist{\'o}bal and de Castro, Pablo and Soto, Rodrigo},
  journal={Physical Review E},
  volume={101},
  number={6},
  pages={062607},
  year={2020},
  publisher={APS}
}

@article{olsen2024optimal,
  title={Optimal diffusion of chiral active particles with strategic reorientations},
  author={Olsen, Kristian St{\o}levik and L{\"o}wen, Hartmut},
  journal={Physical Review E},
  volume={110},
  number={6},
  pages={064606},
  year={2024},
  publisher={APS}
}

@article{potomkin2017focusing,
  title={Focusing of active particles in a converging flow},
  author={Potomkin, Mykhailo and Kaiser, Andreas and Berlyand, Leonid and Aranson, Igor},
  journal={New Journal of Physics},
  volume={19},
  number={11},
  pages={115005},
  year={2017},
  publisher={IOP Publishing}
}

@article{grossmann2016diffusion,
  title={Diffusion properties of active particles with directional reversal},
  author={Gro{\ss}mann, Robert and Peruani, Fernando and B{\"a}r, Markus},
  journal={New Journal of Physics},
  volume={18},
  number={4},
  pages={043009},
  year={2016},
  publisher={IOP Publishing}
}

@article{othmer1988models,
  title={Models of dispersal in biological systems},
  author={Othmer, Hans G and Dunbar, Steven R and Alt, Wolfgang},
  journal={Journal of mathematical biology},
  volume={26},
  number={3},
  pages={263--298},
  year={1988},
  publisher={Springer}
}

@article{hill2007hydrodynamic,
  title={Hydrodynamic Surface Interactions Enable {E}scherichia {C}oli to Seek Efficient Routes to Swim Upstream},
  author={Hill, Jane and Kalkanci, Ozge and McMurry, Jonathan L and Koser, Hur},
  journal={Physical review letters},
  volume={98},
  number={6},
  pages={068101},
  year={2007},
  publisher={APS}
}

@article{marcos2012bacterial,
  title={Bacterial rheotaxis},
  author={Marcos and Fu, Henry C and Powers, Thomas R and Stocker, Roman},
  journal={Proceedings of the National Academy of Sciences},
  volume={109},
  number={13},
  pages={4780--4785},
  year={2012},
  publisher={National Academy of Sciences}
}

@article{palacci2015artificial,
  title={Artificial rheotaxis},
  author={Palacci, J{\'e}r{\'e}mie and Sacanna, Stefano and Abramian, Ana{\"\i}s and Barral, J{\'e}r{\'e}mie and Hanson, Kasey and Grosberg, Alexander Y and Pine, David J and Chaikin, Paul M},
  journal={Science advances},
  volume={1},
  number={4},
  pages={e1400214},
  year={2015},
  publisher={American Association for the Advancement of Science}
}

@article{yuan2015hydrodynamic,
  title={A hydrodynamic mechanism for attraction of undulatory microswimmers to surfaces (bordertaxis)},
  author={Yuan, Jinzhou and Raizen, David M and Bau, Haim H},
  journal={Journal of The Royal Society Interface},
  volume={12},
  number={109},
  pages={20150227},
  year={2015},
  publisher={The Royal Society}
}

@article{ren2017rheotaxis,
  title={Rheotaxis of bimetallic micromotors driven by chemical--acoustic hybrid power},
  author={Ren, Liqiang and Zhou, Dekai and Mao, Zhangming and Xu, Pengtao and Huang, Tony Jun and Mallouk, Thomas E},
  journal={ACS nano},
  volume={11},
  number={10},
  pages={10591--10598},
  year={2017},
  publisher={ACS Publications}
}

@article{figueroa2015living,
  title={Living on the edge: transfer and traffic of {E}. coli in a confined flow},
  author={Figueroa-Morales, Nuris and Mi{\~n}o, Gast{\'o}n Leonardo and Rivera, Aramis and Caballero, Rogelio and Cl{\'e}ment, Eric and Altshuler, Ernesto and Lindner, Anke},
  journal={Soft matter},
  volume={11},
  number={31},
  pages={6284--6293},
  year={2015},
  publisher={Royal Society of Chemistry}
}

@article{pietrangeli2024optimal,
  title={Optimal run-and-tumble in slit-like confinement},
  author={Pietrangeli, T and Ybert, C and Cottin-Bizonne, C and Detcheverry, F},
  journal={Physical Review Research},
  volume={6},
  number={2},
  pages={023028},
  year={2024},
  publisher={APS}
}

@article{codling2008random,
  title={Random walk models in biology},
  author={Codling, Edward A and Plank, Michael J and Benhamou, Simon},
  journal={Journal of the Royal society interface},
  volume={5},
  number={25},
  pages={813--834},
  year={2008},
  publisher={The Royal Society London}
}

@article{malakar2018steady,
  title={Steady state, relaxation and first-passage properties of a run-and-tumble particle in one-dimension},
  author={Malakar, Kanaya and Jemseena, V and Kundu, Anupam and Kumar, K Vijay and Sabhapandit, Sanjib and Majumdar, Satya N and Redner, S and Dhar, Abhishek},
  journal={Journal of Statistical Mechanics: Theory and Experiment},
  volume={2018},
  number={4},
  pages={043215},
  year={2018},
  publisher={IOP Publishing}
}

@article{galajda2007wall,
  title={A wall of funnels concentrates swimming bacteria},
  author={Galajda, Peter and Keymer, Juan and Chaikin, Paul and Austin, Robert},
  journal={Journal of bacteriology},
  volume={189},
  number={23},
  pages={8704--8707},
  year={2007},
  publisher={American Society for Microbiology}
}

@article{zaburdaev2015levy,
  title={L{\'e}vy walks},
  author={Zaburdaev, Vasily and Denisov, Sergey and Klafter, Joseph},
  journal={Reviews of Modern Physics},
  volume={87},
  number={2},
  pages={483--530},
  year={2015},
  publisher={APS}
}

@book{berg2004coli,
  title={E. coli in Motion},
  author={Berg, Howard C},
  year={2004},
  publisher={Springer}
}

@article{korobkova2004molecular,
  title={From molecular noise to behavioural variability in a single bacterium},
  author={Korobkova, Ekaterina and Emonet, Thierry and Vilar, Jose MG and Shimizu, Thomas S and Cluzel, Philippe},
  journal={Nature},
  volume={428},
  number={6982},
  pages={574--578},
  year={2004},
  publisher={Nature Publishing Group UK London}
}

@article{drescher2011fluid,
  title={Fluid dynamics and noise in bacterial cell--cell and cell--surface scattering},
  author={Drescher, Knut and Dunkel, J{\"o}rn and Cisneros, Luis H and Ganguly, Sujoy and Goldstein, Raymond E},
  journal={Proceedings of the National Academy of Sciences},
  volume={108},
  number={27},
  pages={10940--10945},
  year={2011},
  publisher={National Academy of Sciences}
}

@article{nicolle2014catheter,
  title={Catheter associated urinary tract infections},
  author={Nicolle, Lindsay E},
  journal={Antimicrobial resistance and infection control},
  volume={3},
  pages={1--8},
  year={2014},
  publisher={Springer}
}

@article{wright2005uropathogenic,
  title={Uropathogenic Escherichia coli flagella aid in efficient urinary tract colonization},
  author={Wright, Kelly J and Seed, Patrick C and Hultgren, Scott J},
  journal={Infection and immunity},
  volume={73},
  number={11},
  pages={7657--7668},
  year={2005},
  publisher={American Society for Microbiology}
}

@article{dohnt2011vitro,
  title={An in vitro urinary tract catheter system to investigate biofilm development in catheter-associated urinary tract infections},
  author={Dohnt, Katrin and Sauer, Marie and M{\"u}ller, Maren and Atallah, Karin and Weidemann, Marina and Gronemeyer, Petra and Rasch, Detlev and Tielen, Petra and Krull, Rainer},
  journal={Journal of microbiological methods},
  volume={87},
  number={3},
  pages={302--308},
  year={2011},
  publisher={Elsevier}
}

@article{marchetti2013hydrodynamics,
  title={Hydrodynamics of soft active matter},
  author={Marchetti, M Cristina and Joanny, Jean-Fran{\c{c}}ois and Ramaswamy, Sriram and Liverpool, Tanniemola B and Prost, Jacques and Rao, Madan and Simha, R Aditi},
  journal={Reviews of modern physics},
  volume={85},
  number={3},
  pages={1143--1189},
  year={2013},
  publisher={APS}
}

@article{dombrowski2004self,
  title={Self-concentration and large-scale coherence in bacterial dynamics},
  author={Dombrowski, Christopher and Cisneros, Luis and Chatkaew, Sunita and Goldstein, Raymond E and Kessler, John O},
  journal={Physical review letters},
  volume={93},
  number={9},
  pages={098103},
  year={2004},
  publisher={APS}
}

@article{berg1972chemotaxis,
  title={Chemotaxis in Escherichia coli analysed by three-dimensional tracking},
  author={Berg, Howard C and Brown, Douglas A},
  journal={nature},
  volume={239},
  number={5374},
  pages={500--504},
  year={1972},
  publisher={Nature Publishing Group UK London}
}

@book{berg1993random,
  title={Random walks in biology},
  author={Berg, Howard C},
  year={1993},
  publisher={Princeton University Press}
}

@article{sneddon2012stochastic,
  title={Stochastic coordination of multiple actuators reduces latency and improves chemotactic response in bacteria},
  author={Sneddon, Michael W and Pontius, William and Emonet, Thierry},
  journal={Proceedings of the National Academy of Sciences},
  volume={109},
  number={3},
  pages={805--810},
  year={2012},
  publisher={National Academy of Sciences}
}

@article{figueroa20203d,
  title={3{D} spatial exploration by {E}. coli echoes motor temporal variability},
  author={Figueroa-Morales, Nuris and Soto, Rodrigo and Junot, Gaspard and Darnige, Thierry and Douarche, Carine and Martinez, Vincent A and Lindner, Anke and Cl{\'e}ment, {\'E}ric},
  journal={Physical Review X},
  volume={10},
  number={2},
  pages={021004},
  year={2020},
  publisher={APS}
}

@article{matthaus2011origin,
  title={On the origin and characteristics of noise-induced {L}{\'e}vy walks of {E}. coli},
  author={Matth{\"a}us, Franziska and Mommer, Mario S and Curk, Tine and Dobnikar, Jure},
  journal={PloS one},
  volume={6},
  number={4},
  pages={e18623},
  year={2011},
  publisher={Public Library of Science San Francisco, USA}
}

@article{bukatin2020self,
  title={Self-organization and multi-line transport of human spermatozoa in rectangular microchannels due to cell-cell interactions},
  author={Bukatin, A and Denissenko, P and Kantsler, V},
  journal={Scientific reports},
  volume={10},
  number={1},
  pages={9830},
  year={2020},
  publisher={Nature Publishing Group UK London}
}

@article{rubio2021self,
  title={Self-propulsion and shear flow align active particles in nozzles and channels},
  author={Rubio, Leonardo Dominguez and Potomkin, Mykhailo and Baker, Remmi Danae and Sen, Ayusman and Berlyand, Leonid and Aranson, Igor S},
  journal={Advanced Intelligent Systems},
  volume={3},
  number={2},
  pages={2000178},
  year={2021},
  publisher={Wiley Online Library}
}

@article{ezhilan2015transport,
  title={Transport of a dilute active suspension in pressure-driven channel flow},
  author={Ezhilan, Barath and Saintillan, David},
  journal={Journal of Fluid Mechanics},
  volume={777},
  pages={482--522},
  year={2015},
  publisher={Cambridge University Press}
}

@article{bar2020self,
  title={Self-propelled rods: Insights and perspectives for active matter},
  author={B{\"a}r, Markus and Gro{\ss}mann, Robert and Heidenreich, Sebastian and Peruani, Fernando},
  journal={Annual Review of Condensed Matter Physics},
  volume={11},
  number={1},
  pages={441--466},
  year={2020},
  publisher={Annual Reviews}
}

@article{hallatschek2023proliferating,
  title={Proliferating active matter},
  author={Hallatschek, Oskar and Datta, Sujit S and Drescher, Knut and Dunkel, J{\"o}rn and Elgeti, Jens and Waclaw, Bartek and Wingreen, Ned S},
  journal={Nature Reviews Physics},
  volume={5},
  number={7},
  pages={407--419},
  year={2023},
  publisher={Nature Publishing Group UK London}
}

@article{sokolov2009reduction,
  title={Reduction of viscosity in suspension of swimming bacteria},
  author={Sokolov, Andrey and Aranson, Igor S},
  journal={Physical review letters},
  volume={103},
  number={14},
  pages={148101},
  year={2009},
  publisher={APS}
}

@article{haines2009three,
  title={Three-dimensional model for the effective viscosity of bacterial suspensions},
  author={Haines, Brian M and Sokolov, Andrey and Aranson, Igor S and Berlyand, Leonid and Karpeev, Dmitry A},
  journal={Physical Review E—Statistical, Nonlinear, and Soft Matter Physics},
  volume={80},
  number={4},
  pages={041922},
  year={2009},
  publisher={APS}
}

@article{ryan2011viscosity,
  title={Viscosity of bacterial suspensions: Hydrodynamic interactions and self-induced noise},
  author={Ryan, Shawn D and Haines, Brian M and Berlyand, Leonid and Ziebert, Falko and Aranson, Igor S},
  journal={Physical Review E—Statistical, Nonlinear, and Soft Matter Physics},
  volume={83},
  number={5},
  pages={050904},
  year={2011},
  publisher={APS}
}

@article{potomkin2016effective,
  title={Effective rheological properties in semi-dilute bacterial suspensions},
  author={Potomkin, Mykhailo and Ryan, Shawn D and Berlyand, Leonid},
  journal={Bulletin of mathematical biology},
  volume={78},
  pages={580--615},
  year={2016},
  publisher={Springer}
}

@article{paxton2005motility,
  title={Motility of catalytic nanoparticles through self-generated forces},
  author={Paxton, Walter F and Sen, Ayusman and Mallouk, Thomas E},
  journal={Chemistry--A European Journal},
  volume={11},
  number={22},
  pages={6462--6470},
  year={2005},
  publisher={Wiley Online Library}
}

@article{paxton2006chemical,
  title={Chemical locomotion},
  author={Paxton, Walter F and Sundararajan, Shakuntala and Mallouk, Thomas E and Sen, Ayusman},
  journal={Angewandte Chemie International Edition},
  volume={45},
  number={33},
  pages={5420--5429},
  year={2006},
  publisher={Wiley Online Library}
}

@article{rothschild1963non,
  title={Non-random distribution of bull spermatozoa in a drop of sperm suspension},
  author={Rothschild},
  journal={Nature},
  volume={198},
  number={4886},
  pages={1221--1222},
  year={1963},
  publisher={Nature Publishing Group UK London}
}

@article{tailleur2008statistical,
  title={Statistical mechanics of interacting run-and-tumble bacteria},
  author={Tailleur, Julien and Cates, Michael E},
  journal={Physical review letters},
  volume={100},
  number={21},
  pages={218103},
  year={2008},
  publisher={APS}
}

@article{schnitzer1993theory,
  title={Theory of continuum random walks and application to chemotaxis},
  author={Schnitzer, Mark J},
  journal={Physical Review E},
  volume={48},
  number={4},
  pages={2553},
  year={1993},
  publisher={APS}
}

@article{othmer2000diffusion,
  title={The diffusion limit of transport equations derived from velocity-jump processes},
  author={Othmer, Hans G and Hillen, Thomas},
  journal={SIAM Journal on Applied Mathematics},
  volume={61},
  number={3},
  pages={751--775},
  year={2000},
  publisher={SIAM}
}

@article{erban2004individual,
  title={From individual to collective behavior in bacterial chemotaxis},
  author={Erban, Radek and Othmer, Hans G},
  journal={SIAM Journal on Applied Mathematics},
  volume={65},
  number={2},
  pages={361--391},
  year={2004},
  publisher={SIAM}
}

@book{tabeling2023introduction,
  title={Introduction to microfluidics},
  author={Tabeling, Patrick},
  year={2023},
  publisher={Oxford university press}
}

@book{weiss1994aspects,
  title={Aspects and applications of the random walk},
  author={Weiss, George Herbert},
  publisher={North-Holland, New York},
  year={1994}
}

@article{masoliver2017continuous,
  title={Continuous time persistent random walk: a review and some generalizations},
  author={Masoliver, Jaume and Lindenberg, Katja},
  journal={The European Physical Journal B},
  volume={90},
  number={6},
  pages={107},
  year={2017},
  publisher={Springer}
}

@article{metzler2000random,
  title={The random walk's guide to anomalous diffusion: a fractional dynamics approach},
  author={Metzler, Ralf and Klafter, Joseph},
  journal={Physics reports},
  volume={339},
  number={1},
  pages={1--77},
  year={2000},
  publisher={Elsevier}
}

@article{othmer2002diffusion,
  title={The diffusion limit of transport equations {I}{I}: Chemotaxis equations},
  author={Othmer, Hans G and Hillen, Thomas},
  journal={SIAM Journal on Applied Mathematics},
  volume={62},
  number={4},
  pages={1222--1250},
  year={2002},
  publisher={SIAM}
}

@article{montroll1965random,
  title={Random walks on lattices. {I}{I}},
  author={Montroll, Elliott W and Weiss, George H},
  journal={Journal of Mathematical Physics},
  volume={6},
  number={2},
  pages={167--181},
  year={1965},
  publisher={American Institute of Physics}
}

@incollection{risken1989fokker,
  title={Fokker-{P}lanck equation},
  author={Risken, Hannes},
  booktitle={The Fokker-Planck equation: methods of solution and applications},
  pages={63--95},
  year={1989},
  publisher={Springer}
}

@article{ariel2015swarming,
  title={Swarming bacteria migrate by {L}{\'e}vy Walk},
  author={Ariel, Gil and Rabani, Amit and Benisty, Sivan and Partridge, Jonathan D and Harshey, Rasika M and Be'er, Avraham},
  journal={Nature communications},
  volume={6},
  number={1},
  pages={8396},
  year={2015},
  publisher={Nature Publishing Group UK London}
}

@article{berlyand2026multiscale,
  title={Multiscale analysis of a kinetic model of confined suspensions of self-propelled rods},
  author={Berlyand, Leonid and Dang, Spencer and Jabin, Pierre-Emmanuel and Potomkin, Mykhailo},
  journal={Mathematical Models and Methods in Applied Sciences},
  volume={36},
  number={03},
  pages={629--654},
  year={2026},
  publisher={World Scientific}
}

@article{jeffery1922motion,
  title={The motion of ellipsoidal particles immersed in a viscous fluid},
  author={Jeffery, George Barker},
  journal={Proceedings of the Royal Society of London. Series A, Containing papers of a mathematical and physical character},
  volume={102},
  number={715},
  pages={161--179},
  year={1922},
  publisher={The Royal Society London}
}

@article{karrila1992microhydrodynamics,
  title={Microhydrodynamics: principles and selected applications},
  author={Kim, Sagtae and Karrila, Seppo},
  journal={Applied Mathematical Modelling},
  year={1992}
}

@article{zhou2024ai,
  title={AI-aided geometric design of anti-infection catheters},
  author={Zhou, Tingtao and Wan, Xuan and Huang, Daniel Zhengyu and Li, Zongyi and Peng, Zhiwei and Anandkumar, Anima and Brady, John F and Sternberg, Paul W and Daraio, Chiara},
  journal={Science Advances},
  volume={10},
  number={1},
  pages={eadj1741},
  year={2024},
  publisher={American Association for the Advancement of Science}
}

@article{tao2026invasion,
  title={Invasion of bacteria swimming upstream into microstructured devices},
  author={Tao, Ran and Th{\'e}ry, Albane and Que, Suya and Mathijssen, Arnold JTM},
  journal={Newton},
  volume={2},
  number={1},
  year={2026},
  publisher={Elsevier}
}

@article{peng2020upstream,
  title={Upstream swimming and Taylor dispersion of active Brownian particles},
  author={Peng, Zhiwei and Brady, John F},
  journal={Physical Review Fluids},
  volume={5},
  number={7},
  pages={073102},
  year={2020},
  publisher={APS}
}

\end{document}